\documentclass[]{spie}  

\usepackage{amsmath,amsfonts,amssymb}
\usepackage{graphicx}
\usepackage{siunitx}
\usepackage[colorlinks=true, allcolors=blue]{hyperref}
\usepackage[dvipsnames]{xcolor}
\usepackage{textgreek}

\newcommand{\LL}[1]{\textcolor{black}{#1}}
\newcommand{\AR}[1]{\textcolor{black}{#1}}
\newcommand{\JEG}[1]{\textcolor{black}{#1}}

\newcommand{\CF}[1]{\textcolor{black}{#1}}

\title{The optical design of the LiteBIRD Middle and High Frequency Telescope}

\author[a,r]{L.~Lamagna}
\author[b]{J.E.~Gudmundsson}
\author[c]{H.~Imada}
\author[d]{P.~Hargrave}
\author[e,s]{C.~Franceschet}
\author[a,r]{M.~De~Petris}
\author[f]{J.~Austermann}
\author[g]{S.~Bounissou}
\author[a,r]{F.~Columbro}
\author[a,r]{P.~de~Bernardis}
\author[h]{S.~Henrot-Versill\'e}
\author[f]{J.~Hubmayr}
\author[i]{G.~Jaehnig}
\author[j]{R.~Keskitalo}
\author[g]{B.~Maffei}
\author[a,r]{S.~Masi}
\author[k]{T.~Matsumura}
\author[l]{L.~Montier}
\author[l]{B.~Mot}
\author[d]{F.~Noviello}
\author[m]{C.~O'Sullivan}
\author[a,r]{A.~Paiella}
\author[a]{G.~Pisano}
\author[e,s]{S.~Realini}
\author[g,q]{A.~Ritacco}
\author[n]{G.~Savini}
\author[o]{A.~Suzuki}
\author[m]{N.~Trappe}
\author[t]{B.~Winter\textsuperscript{p}, for the Litebird Joint Study Group}

\affil[a]{Physics Department, Sapienza University of Rome, Italy}
\affil[b]{The Oskar Klein Centre, Department of Physics, Stockholm University, Sweden}
\affil[c]{National Astronomical Observatory of Japan, Osawa, Mitaka, Japan}
\affil[d]{School of Physics and Astronomy, Cardiff University, UK}
\affil[e]{Physics Department, Università degli Studi di Milano Statale, Italy}
\affil[f]{National Institute of Standards and Technology, Boulder, Colorado, US}
\affil[g]{Institut d'Astrophysique Spatiale, Université Paris-Saclay, France}
\affil[h]{Universit\'e Paris-Saclay, CNRS/IN2P3, IJCLab, Orsay, France}
\affil[i]{Center for Astrophysics and Space Astronomy, University of Colorado Boulder, US}
\affil[j]{Computational Cosmology Center, Lawrence Berkeley National Lab, US}
\affil[k]{Kavli Institute for the Physics and Mathematics of the Universe, Kashiwa, Japan}
\affil[l]{Institut de Recherche en Astrophysique et Planétologie, Toulouse, France}
\affil[m]{Department of Experimental Physics, Maynooth University, Ireland}
\affil[n]{Department of Physics and Astronomy, University College London, UK}
\affil[o]{Physics Division, Lawrence Berkeley National Lab, US}
\affil[p]{Department of Space and Climate Physics, University College London, UK}
\affil[q]{Département de Physique, Ecole Normale Supérieure, Paris, France}
\affil[r]{INFN Sezione di Roma1, Italy}
\affil[s]{INFN Sezione di Milano, Italy}
\affil[t]{LiteBIRD Joint Study Group Members are listed at \href{https://wiki.kek.jp/display/cmb/LiteBIRD+Joint+Study+Group+members+picture+book}{this link}}
\authorinfo{Corresponding author information:\\
E-mail: luca.lamagna@uniroma1.it, Telephone: +39 (0)6 4991 4287}

\begin{document} 
\sisetup{range-phrase=--}
\maketitle

\begin{abstract}
LiteBIRD is a JAXA strategic L-class mission devoted to the measurement of polarization of the Cosmic Microwave Background, searching for the signature of primordial gravitational waves in the B-modes pattern of the polarization. The onboard instrumentation includes a Middle and High Frequency Telescope (MHFT), based on a pair of cryogenically cooled refractive telescopes covering, respectively, the 89-224 GHz and the 166-448 GHz bands. Given the high target sensitivity and the careful systematics control needed to achieve the scientific goals of the mission, optical modeling and characterization are performed with the aim to capture most of the physical effects potentially affecting the real performance of the two refractors.
We describe the main features of the MHFT, its design drivers and the major challenges in system optimization and characterization. We provide the current status of the development of the optical system and we describe the current plan of activities related to optical performance simulation and validation.\end{abstract}



\section{INTRODUCTION}
\label{sec:intro}

LiteBIRD is a planned JAXA satellite which will be inserted into a L2 Lissajous orbit by the end of 2020s with a JAXA H3 rocket, for three years of observations.\cite{JLTPLB} The main aim of this experiment is to detect or constrain the amplitude of a hypothesized primordial $B$-mode signal in the Cosmic Microwave Background (CMB). LiteBIRD comprises two telescopes, the Low Frequency Telescope (LFT), with a frequency range of \mbox{34-\SI{161}{\giga\hertz}} and the Middle and High Frequency Telescope (MHFT) operating in the \mbox{89-\SI{448}{\giga\hertz}} range. Together, the two instruments will observe the sky in 15 spectral bands. This proceeding is dedicated to the optical design of the MHFT. We will review the baseline design and discuss initial results of various optical simulations. 

\section{Middle and High Frequency Telescope Design}
\label{sec:mhftdesc}
\subsection{Choice of refractive over reflective design}
\begin{figure} [ht]
   \begin{center}
   \begin{tabular}{c} 
   \includegraphics[height=7cm]{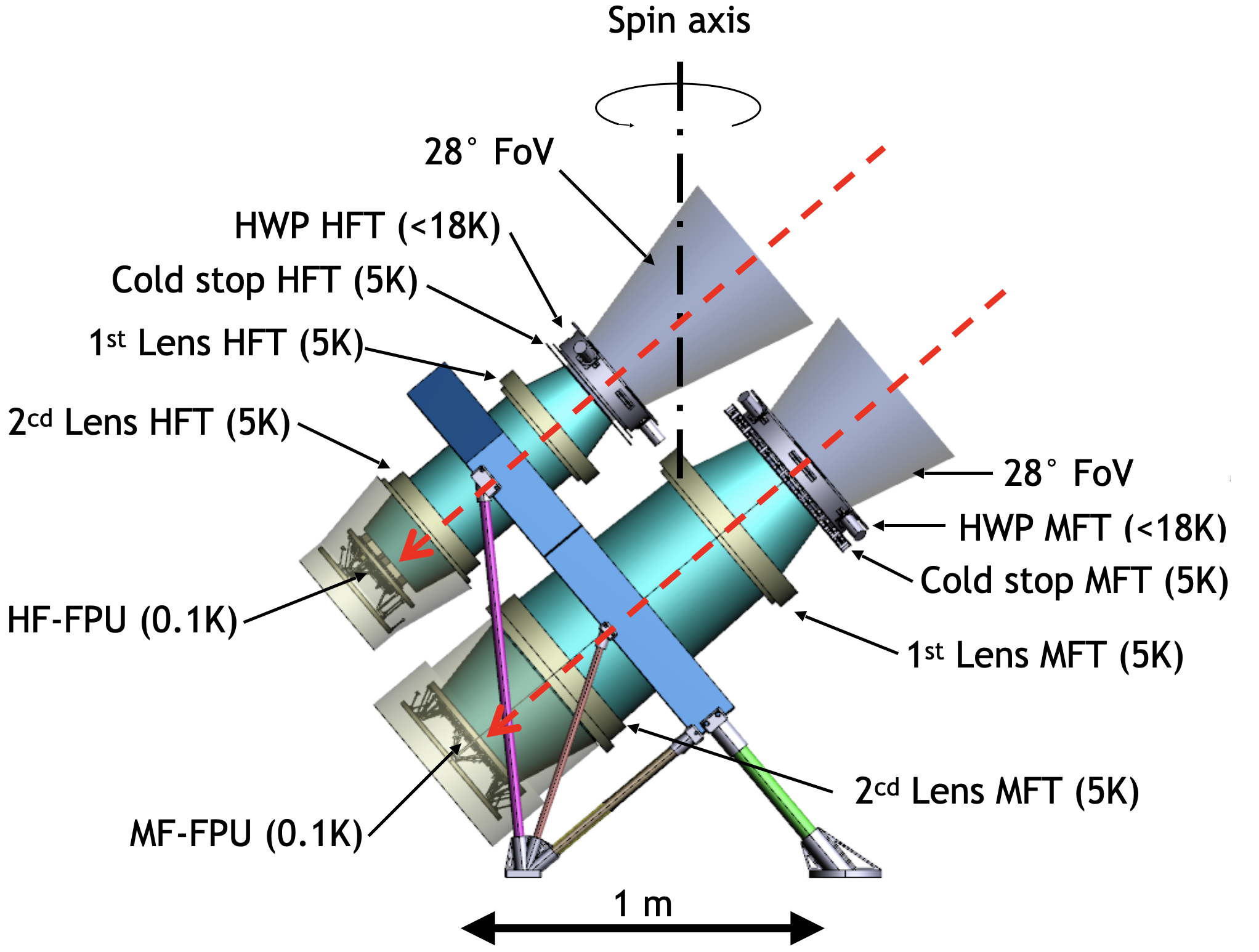}
   \end{tabular}
   \end{center}
   \caption[example] 
   { \label{fig:Mechanical_design-optionB} 
MHFT overview, the various subsystems composing the telescopes are identified, and held by the mechanical structure. The optical front baffles still have to be designed.}
   \end{figure}

Polarization modulation affected the MHFT design from its earliest development phases.
In fact, the MHFT trade-off analysis has been mainly driven by the constraints due to the Half Wave Plate (HWP) technology. The wide frequency coverage, from 89 GHz to 448 GHz, associated to the limited mass budget placed strong constraints on the HWP material. The weight of a multi-layer sapphire HWP, and the difficulty in applying an anti-reflection coating (ARC) for the highest frequency channels led to discard this solution in favour of a metamaterial-based design (mHWP)\cite{Pisano2012,Pisano2016,columbro2020,Pisano2020}. 
In the framework of the ESA CDF (Concurrent Design Facility) study, the two following designs have been studied:
	\begin{itemize}
	    \item Option A - fully reflective: single reflective optics + reflective mHWP\cite{Pisano2016} covering the full frequency range.
	    \item Option B - fully refractive: two telescopes with refractive optics + transmissive mHWP\cite{Pisano2012}, by splitting the total frequency range into mid-frequency (MFT) and high-frequency (HFT) ranges.
	\end{itemize}
Following the end of the ESA-CDF study, further work and optimizations have been done on the design of both option A \& B and most of the issues identified during the CDF have been addressed.
Among the possible implementations of mHWP technology, the safest and less prone to unknown optical systematics was determined to be a transmissive, mesh-based multilayer design, where the incoming polarization is rotated while being transmitted through a rotating flat reactive structure, located along the optical path from the sky to the telescope. 
Given the broad spectral coverage, a separation of the MHFT into two distinct optical assemblies (MFT, \mbox{89-\SI{224}{\giga\hertz}} and HFT, \mbox{166-\SI{448}{\giga\hertz}}) has been proposed, each with its own polarization modulator, in order to mitigate the bandwidth issues of the mHWP technology. 
In combination with mass/volume considerations for the telescope assemblies, and other major system-level tradeoff considerations, this study brought to the choice of the fully refractive, on-axis dual-lens solution as a baseline for both the MFT and for the HFT (see Fig.~\ref{fig:Mechanical_design-optionB}). 

In this configuration, the continuously rotating polarization modulator is the first optical element of the chain, and two AR-coated plastic lenses in a telecentric configuration focus radiation on a flat focal surface where the Transition Edge Sensor (TES) arrays are accommodated. The aperture stop of the system is located skywards of the objective lens, and the HWP is placed skywards of the aperture stop, just as close to it as practically allowed by mechanical and thermal considerations, in order to optimize the size and mass of the PMU mechanism. Quasi-optical filters, based on plastic-embedded metal-mesh technology, are placed along the optical chain to prevent far-infrared radiation from propagating all the way to the focal plane, controlling the overall power loading on the detectors. Microwave absorbing layers on the tube walls and around the aperture stop are implemented to ensure further control of in-band stray-light. Final frequency selection and band refinement for each subset of detectors take place on the focal plane assemblies. The whole telescope unit, including the tube, the PMU, the lenses and the filters, is cooled down to 4.8\,K. A 1.8\,K cold baffle surrounds the 100\, mK Focal Plane Unit (FPU), to provide both additional loading control for the environment surrounding the FPUs and a stage of magnetic shielding for the Transition Edge detectors.
Cold forebaffles at the sky input of the tubes will be designed to ensure reliable control of beam far sidelobes and prevent radiation pickup from the emissive payload structure in close proximity to each telescope.

The major strengths of the fully refractive configuration are:
\begin{enumerate}
\item In addition to being simple and conventional, the design strongly benefits from the broad expertise gained on many current and upcoming sub-orbital CMB experiments, such as BICEP2\cite{BICEP2}, Keck\cite{KECK}, SPIDER\cite{Rahlin2014}, LSPE\cite{LSPE}, and Simons Observatory\cite{SimonsObs}
\item With these very compact on-axis designs, the MFT and HFT telescopes offer the necessary design flexibility to match the volume and mass constraints, to mitigate stray-light issues, and to split the entire frequency range in two bands. 
\item This choice has also obvious simplifications for the design of the separate filtering chains. 
\item Dealing with two smaller compact telescopes will be considerably more practical for the Assembly Integration and Test (AIT) and the Assembly Integration Verification (AIV) phases of the project, and for all the ground calibration activities.
\end{enumerate}

On the other hand, the use of dielectric lenses implies the need for a careful design and modeling, and a dedicated performance validation effort. This is needed to address both the issues related to intrinsic frequency, polarization and temperature-dependent properties of the refracting materials, as well as to understand the higher order optical coupling effects arising from a broad set of non-idealities, such as imperfect dielectric-to-vacuum matching at the optical interfaces (usually compensated through custom, but not ideal, anti-reflection coatings), imperfectly absorbing tube surfaces and focal planes, scattering off internal surfaces, and thermal radiation pickup from the payload environment. Some of these issues are also exacerbated by the modulation from the transmissive HWP: since scattered radiation is likely to be polarized with a well defined but not optimized frequency dependence. If the stray optical paths involve a reflection on the rotating HWP, this will result in a partial modulation of this stray radiation and therefore in a spurious contribution in the signal bandpass, should this radiation propagate back on the detectors.

\subsection{Baseline optical design parameters}\label{sec:optmf}
\begin{figure}[t]
    \centering
    \begin{tabular}{cc}
    \begin{minipage}{0.48\hsize}
    \includegraphics[width = \hsize ]{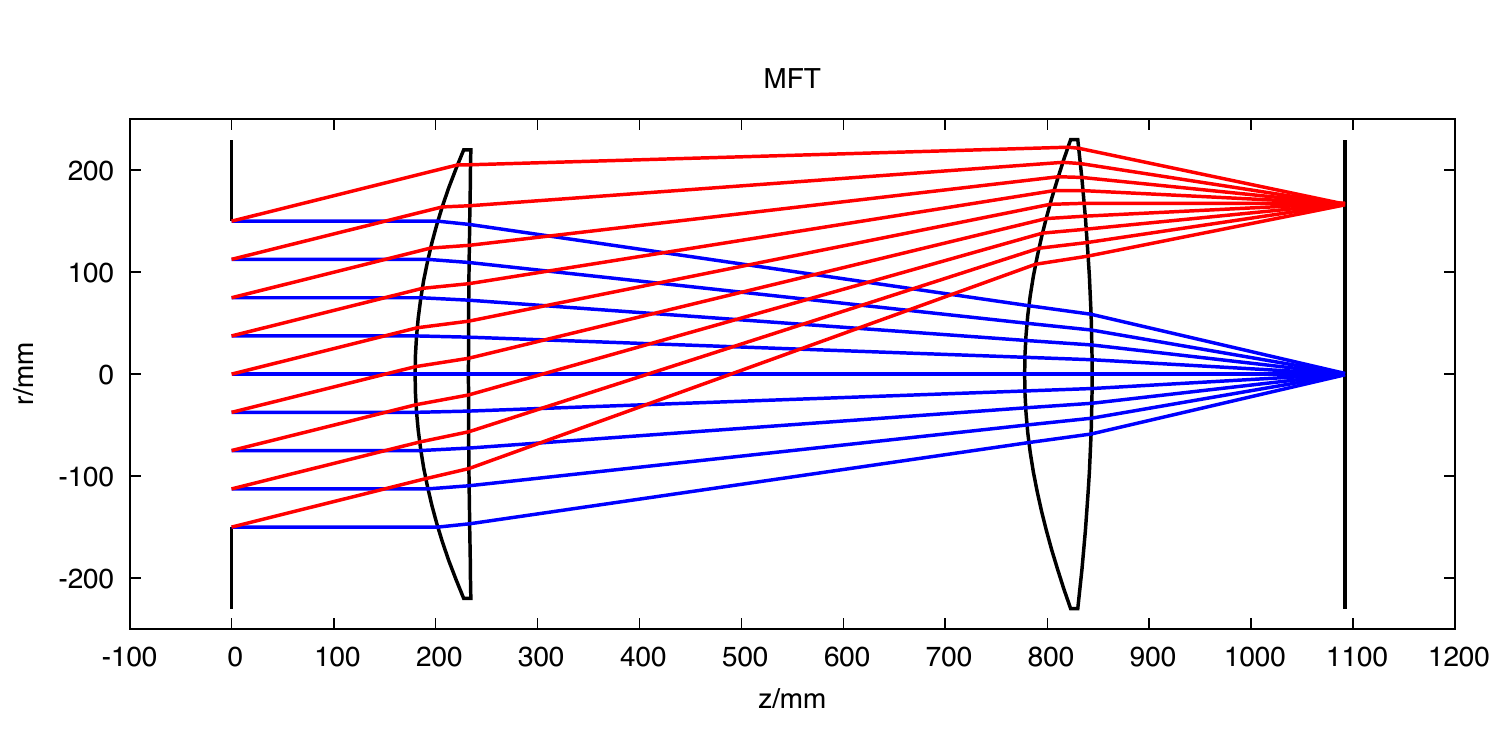}
    \end{minipage} &
    \begin{minipage}{0.48\hsize}
    \includegraphics[width = \hsize ]{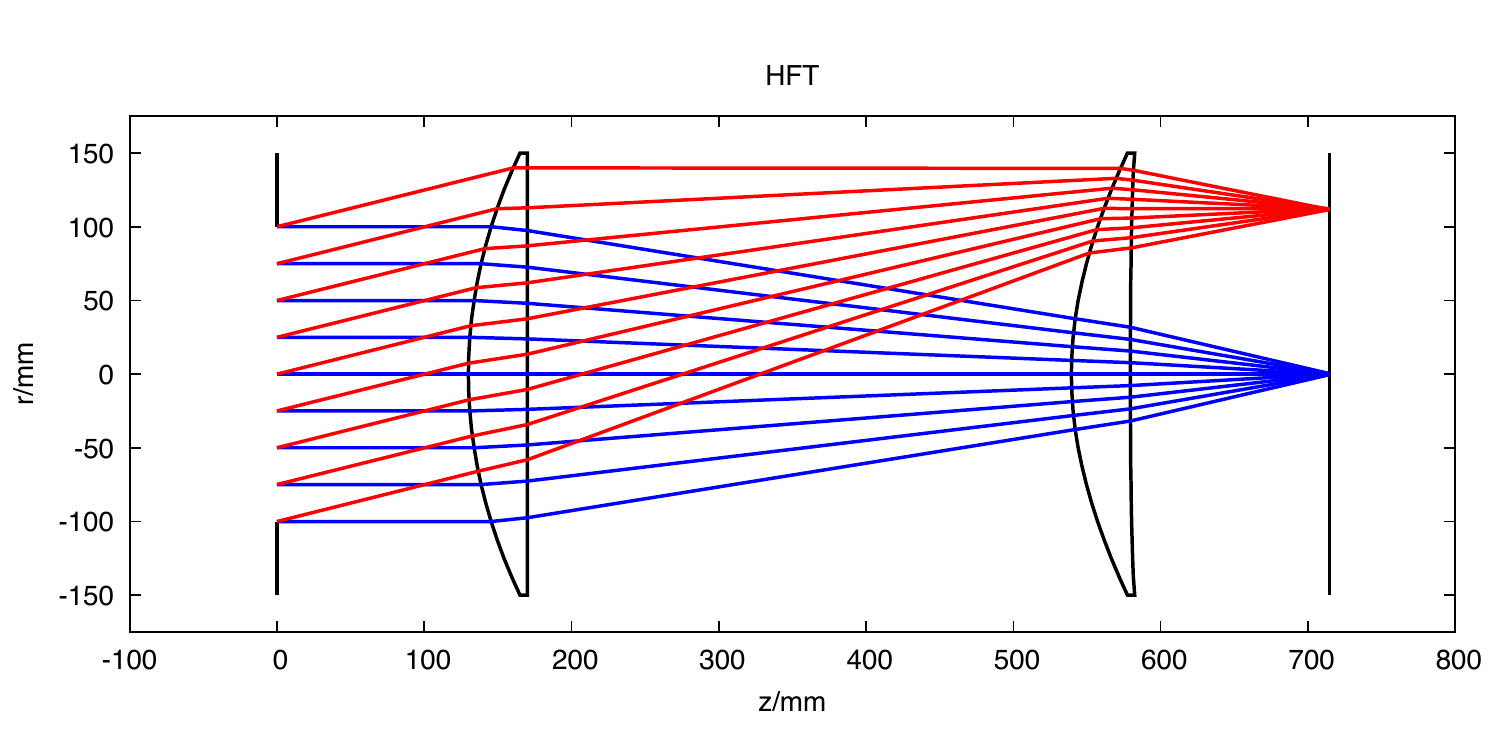}
    \end{minipage}
    \end{tabular}
    \caption{Ray diagrams of MFT (left) and HFT (right). The on-axis and off-axis fields (14 deg) are the blue and red rays, respectively. The telescope aperture is located at $ z = 0 $.}
    \label{fig:opt_config}
\end{figure}

\JEG{
The LiteBIRD MHFT optical designs are based on a simple two-lens refractor concept. Figure \ref{fig:opt_config} shows a ray trace diagram of the two telescope designs. Both telescopes employ two polypropylene (PP) or ultra-high molecular weight polyethylene (UHMW-PE) lenses with an assumed index of refraction $n_\mathrm{r} = 1.52$. The final choice of the plastic material will affect the lens profiles very marginally, even if different high-frequency losses for the plastics may be relevant for HFT.}

\JEG{The lenses will employ one or multi-layer anti-reflection (AR) coatings to support the broad frequency range covered by these telescopes\cite{Pisano2018}. A slight preference for PP lenses emerges in this context, due to the higher melting temperature making the ARC application easier and more reliable.}

\begin{figure}[t]
    \centering
    \begin{tabular}{cc}
    \begin{minipage}{0.48\hsize}
    \includegraphics[width = \hsize ]{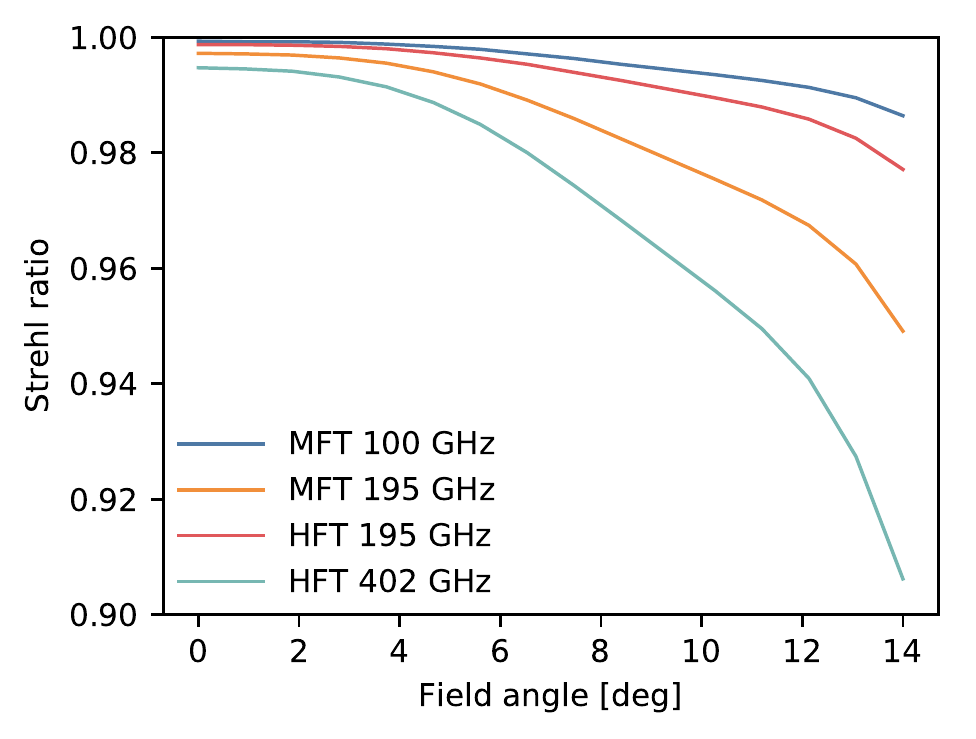}
    \end{minipage} &
    \begin{minipage}{0.48\hsize}
    \includegraphics[width = \hsize ]{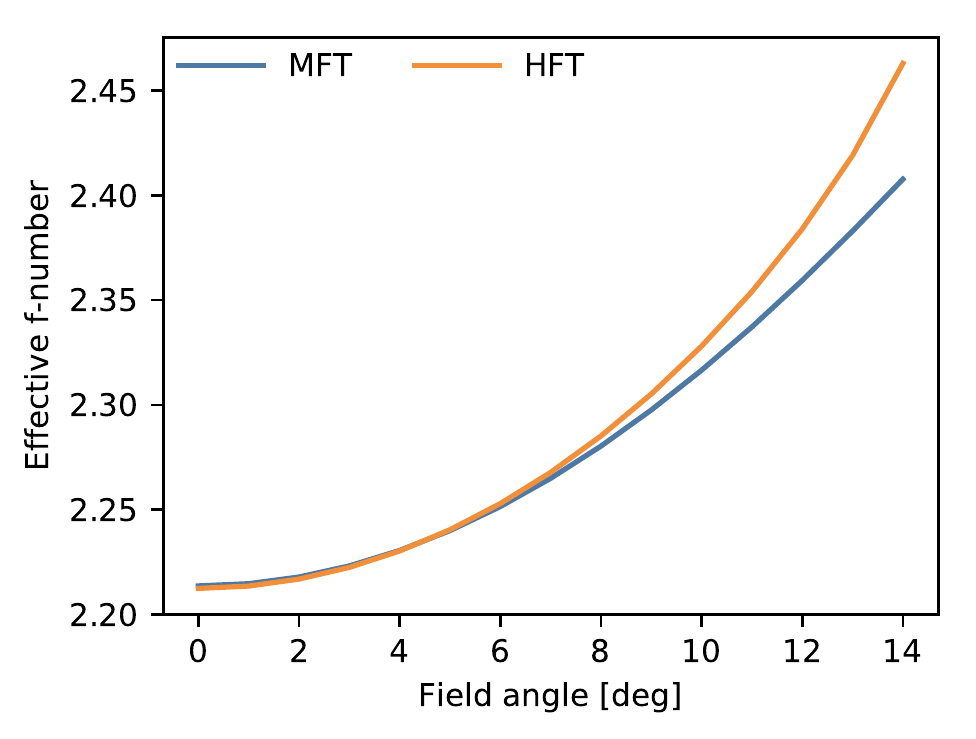}
    \end{minipage}
    \end{tabular}
    \caption{\JEG{Left: Strehl ratio as a function of field angle for the lowest and highest frequency bands on the MFT and HFT. Right: The effective (working) $f$-number variation as a function of field angle for the MFT and HFT.}}
    \label{fig:strehl_fnumber}
\end{figure}

\JEG{The aperture stop is located skywards of the primary lens such that the two telescopes both have an effective (working) $f$-number of 2.2 at the center of the focal plane. The effective $f$-number grows to 2.40-2.45 at the edge of the focal plane. The Strehl ratios and the f-number of the MFT and HFT are shown in Figure \ref{fig:strehl_fnumber}. It is clear that the designs support more than a \ang{28} diffraction-limited field of view.
}
\JEG{
The telescope apertures, which are placed skywards of the primary lens, have diameters of \SI{300}{\milli\metre} and \SI{200}{\milli\metre} for the MFT and HFT, respectively. The designs are fully telecentric with a chief ray angle of incidence not exceeding \ang{0.1} across the field of view. 
}

\section{Preliminary optical modeling results}
Numerical studies are required for the design optimization of MHFT, both because of the complex interplay between different physical aspects of radiation propagation through the different parts of the telescopes, as well as for the broad range of frequencies covered by the two telescopes. Commercially available software tools provide an effective and informative support in this activity, but they often need to be complemented by custom post-processing tools developed for specific studies. In addition, dielectrics with compliant nominal properties are used as lens materials and in the absorptive coatings. Especially at the highest frequencies, these need to be carefully treated in order to capture all the relevant processes affecting the telescope performance, including e.g. dielectric losses and unwanted polarization-sensitive scattering,

Numerical simulations are being performed: 
\begin{itemize}
    \item At the {\it subsystem} level, to refine the design of telescope elements, identify critical tolerances, optimize surface coatings, investigate the broadband behavior of surface shapes and inspect the feasibility of competing solutions for focal plane coupling. 
    \item At the {\it system} level, in order to provide a global picture of the receiver performance, e.g. by delivering performance-relevant quantities like polarization-sensitive sky beams and realistic sidelobe pickup as a function of the detector position on the focal plane. These data products are helpful to inform more accurate TOD-based simulations of the observed microwave sky and to provide inputs for the design and planning of RF--testing and calibration procedures.   
\end{itemize}


\subsection{Preliminary tolerancing results.}
\begin{figure}[tbp]
    \centering
    \begin{minipage}{0.5\hsize}
    \includegraphics[ width = \hsize]{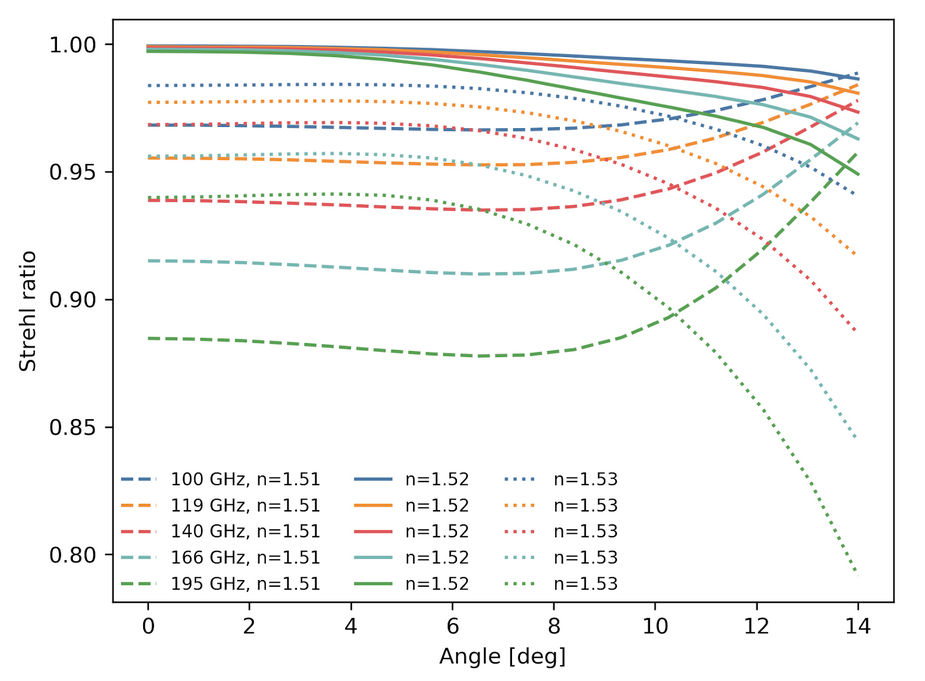}
    \end{minipage} 
    \caption{MFT Strehl ratio across the field of view, as a function of lens refractive index.}
    \label{fig:MFT_SR_vs_n}
\end{figure}

\JEG{Simple geometrical tolerancing has been performed using Zemax. This work suggests that the MHFT design is relatively robust to error in alignment that result in translation and rotation errors with respect to specifications. Out of the tolerancing studies considered so far,  the constraints on lens index of refraction are found to be the most stringent.}
By using the Strehl ratio as a simple performance metric, we find that an uncertainty of 0.01 in the plastic refractive index (assuming a reference value of $n=1.52$), preserves a Strehl ratio greater than 0.8 across the whole field of view of MFT (see Fig.~\ref{fig:MFT_SR_vs_n})
More advanced tolerancing studies are planned for future design iterations. These will include  surface error analyses on different scales, more realistic representation of the tube structures and a dedicated focus on the high-frequency behavior of dielectric materials for the HFT.

\begin{table}[htbp]
    \centering
    \begin{tabular}{l ccccc c ccccc}
        \hline
        & \multicolumn{5}{c}{MFT} & & \multicolumn{5}{c}{HFT} \\
        \cline{2-6} \cline{8-12} 
        & 100 & 119 & 140 & 166 & 195 & & 195 & 238 & 280 & 337 & 402 \\
        \cline{2-6} \cline{8-12} 
        FWHM [arcmin] & 37.8 & 33.4 & 30.4 & 28.3 & 27.2 & & 28.6 & 24.6 & 22.4 & 20.6 & 17.7  \\
        Edge taper [dB] & -9.5 & -13.1 & -17.9 & -24.9 & -34.3 & & -8.4 & -12.1 & -17.1 & -24.6 & -26.1  \\
        Ellipticity & 0.002 & 0.003 & 0.005 & 0.008 & 0.011 & & 0.003 & 0.005 & 0.008 & 0.010 & 0.0115  \\
        \hline\\
    \end{tabular}
    \caption{\label{tab:OpticalParams}Mean predicted optical parameters for the each of the five MFT and HFT frequency bands. The mean is estimated from 12 and 7 simulations for MFT and HFT, respectively (see dicussion in Section \ref{sec:po}).}
\end{table}

\subsection{Baseline physical optics simulation results}
\label{sec:po}
\JEG{Simple physical optics simulations are used to predict the far field response of the proposed optical design. These simulations can be used to study the expected variation in beam size and beam ellipticity as a function of frequency and field location. The same framework can be used to estimate spillover past lenses and the aperture stop (see Section \ref{sec:spillober}). For this work, we make use of a simulation framework first described in Ref.~\citenum{Gudmundsson2020}. We refer the reader to this work for a more detailed discussion of the simulation and analysis methodology. Key optical properties derived from these physical optics simulations are summarized in Table~\ref{tab:OpticalParams}.}

\begin{figure}[htb]
    \centering
    \begin{tabular}{cc}
    \begin{minipage}{0.47\hsize}
    \includegraphics[ width = \hsize]{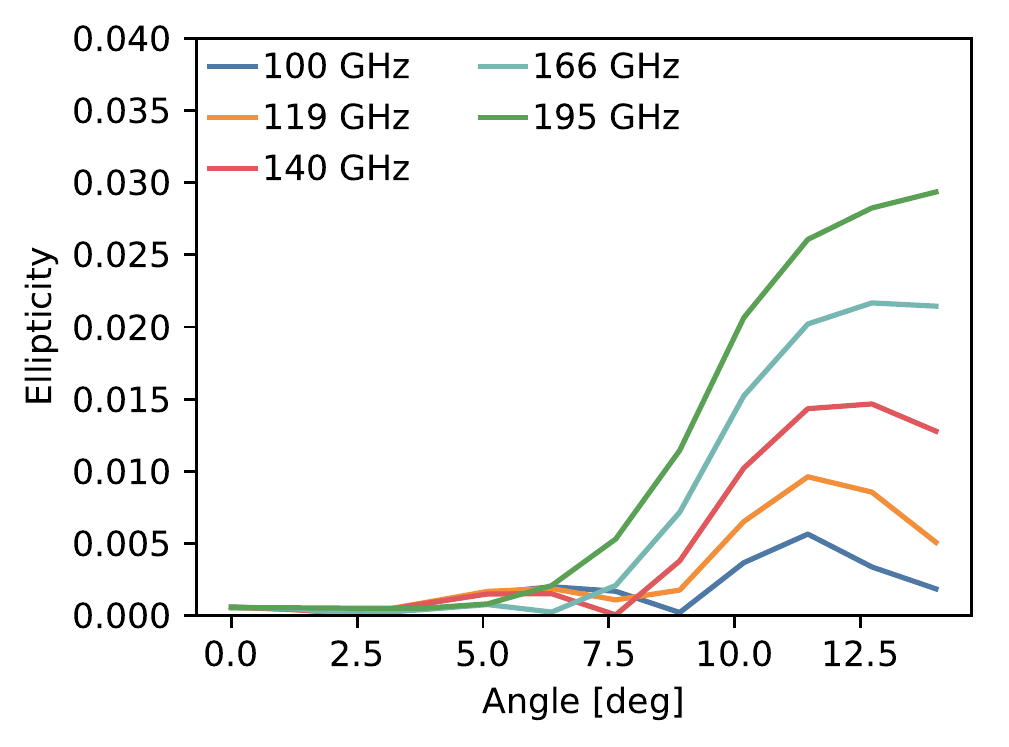}
    \end{minipage} &
    \begin{minipage}{0.47\hsize}
    \includegraphics[ width = \hsize ]{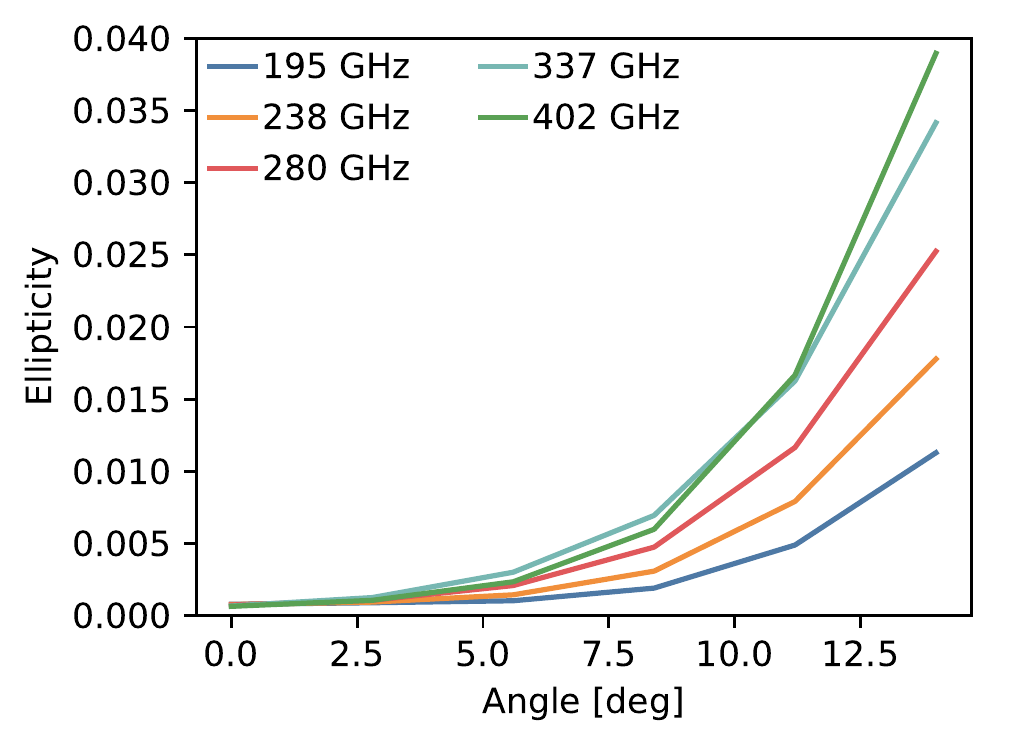}
    \end{minipage}
    \end{tabular}
    \caption{Simulated MFT and HFT beam ellipticity as a function of field angle for all 10 frequency bands.}
    \label{fig:mhft_ellipticity}
\end{figure}

\JEG{The physical optics simulations use Gaussian pixel beams as input. The width of the Gaussian beams correspond to the assumed pixel apertures, 11.6 and \SI{6.6}{\milli\meter} for the MFT and HFT, respectively. In order to optimize  time usage, the simulations are run for a single frequency within a given band. Moreover, because of the azimuthal symmetry of the two-lens refractor design, we only run simulations for a handful of detectors along a radial line from the center of the focal plane out to its edge, this number of detectors being 7 and 12 for the MFT and HFT, respectively. These physical optics simulations involve only the two lenses and the cold aperture stop, also assuming perfect mechanical alignment and material properties while ignoring any reflections. Figure~\ref{fig:mhft_ellipticity} shows the predicted beam ellipticity as a function of field angle for the MFT and HFT frequency bands. We define the beam ellipticity as
\begin{equation}
e = \frac{\sigma_{\mathrm{max}}-\sigma_{\mathrm{min}}}{\sigma_{\mathrm{max}}+\sigma_{\mathrm{min}}},
\end{equation}
where $\sigma _{\mathrm{max}}$ and $\sigma _{\mathrm{min}}$ correspond to the widths of the best fit Gaussian envelope to the semi-major and semi-minor axis of the co-polarized far field beam response.
}

\JEG{As expected, the beam ellipticity is predicted to grow with field location and frequency. We note however that the ellipticity is observed to be quite low even at the edge of the \ang{28} field of view at the highest frequencies. For comparison, the mean beam ellipticity of the 100 and \SI[number-unit-product=\text{-}]{143}{\giga\hertz} channels on Planck was roughly 0.085 and 0.020, respectively \cite{Planck2020_data_processing}.}

\subsection{Spillover simulation results}
\label{sec:spillober}
\JEG{Accurate modeling of optical spillover inside the optics tube is necessary for sensitivity estimates. We can use the physical optics modeling results to estimate the beam power that misses the secondary and primary lenses and the aperture stop. By summing up the optical power that misses these optical elements and comparing it to the pixel beam model input, we can estimate the fractional cold spillover. Further discussion of this simulation approach is described in Ref.~\citenum{Gudmundsson2020}.}

\begin{figure}[t]
    \centering
    \begin{tabular}{cc}
    \begin{minipage}{0.47\hsize}
    \includegraphics[ width = \hsize]{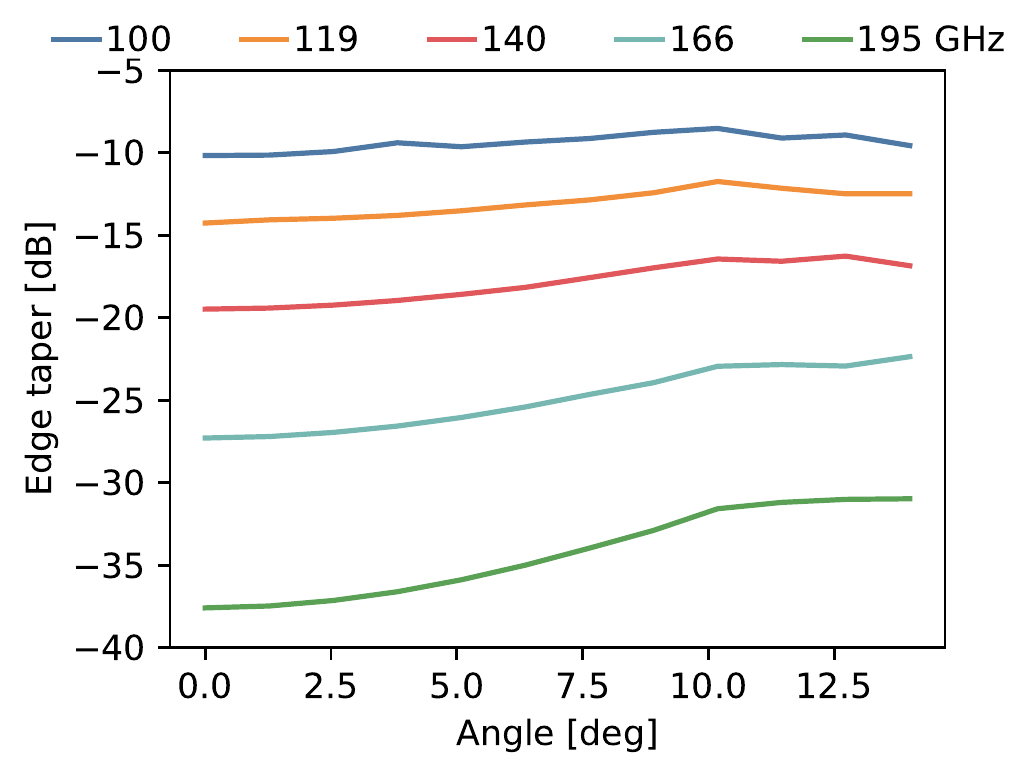}
    \end{minipage} &
    \begin{minipage}{0.47\hsize}
    \includegraphics[ width = \hsize ]{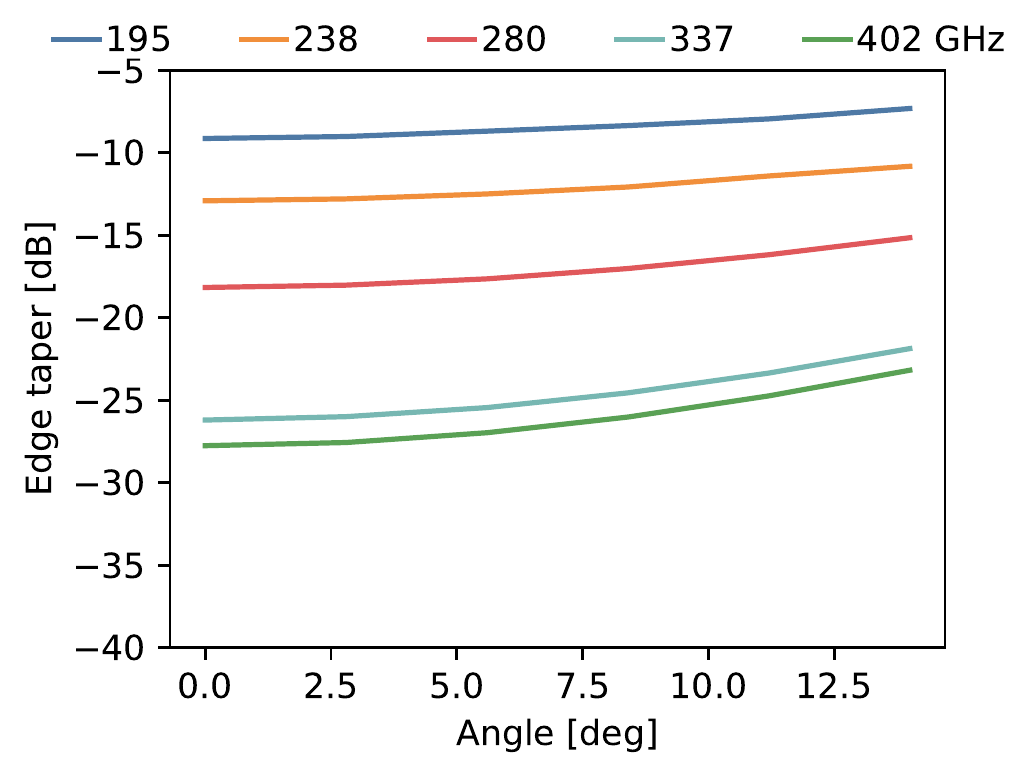}
    \end{minipage}
    \end{tabular}
    \caption{MFT and HFT edge taper as a function of field angle for all 10 frequency bands.}
    \label{fig:mhft_edgetaper}
\end{figure}

\JEG{As the expected pixel beam response of the proposed detector designs was not available, these simulations rely on the same Gaussian pixel beam that is used for analysis described in Section \ref{sec:po}. Figure \ref{fig:mhft_edgetaper} shows the predicted edge taper as a function of focal plane location for the MFT and HFT. For off-axis pixels, we define the edge taper as azimuthally averaged beam power along the stop edge. As expected, we see a slight increase in the edge taper as we move across the focal plane. It is important to note that because of the fixed pixel diameter assumed for the MFT and HFT, we observe significant shift in the edge taper as a function of frequency. Consequently, the conservative edge taper for the high-frequency bands should help reduce sidelobe response.}


\subsection{Inclusion of forebaffle effects}
\CF{In addition to the dual lens optical path response, the final radiation pattern is affected by several elements of the MHFT telescopes – e.g. forebaffle, mHWP, filters, tube with internal absorber, ARC and optical properties of the lenses among the few – each of them having an impact, possibly negligible, on different angular ranges. The most off-axis pixels of the focal plane are characterized by an intrinsic degree of asymmetry which could magnify the effect of these telescope elements on the beam. Therefore, we study the impact of each individual item on the resulting beam separately, in order to assess the angular range which is mainly affected as well as the relevant intensity in terms of sidelobe levels. An analysis to optimize the telescope components is also desirable, as it allows for an estimation of the sensitivity of the beam response to these effects; e.g. the tube diameter, the properties of the tube internal absorber, forebaffle length, etc.}

\begin{figure}[ht]
    \centering
    \begin{tabular}{cc}
    \begin{minipage}{0.5\hsize}
    \includegraphics[width = \hsize]{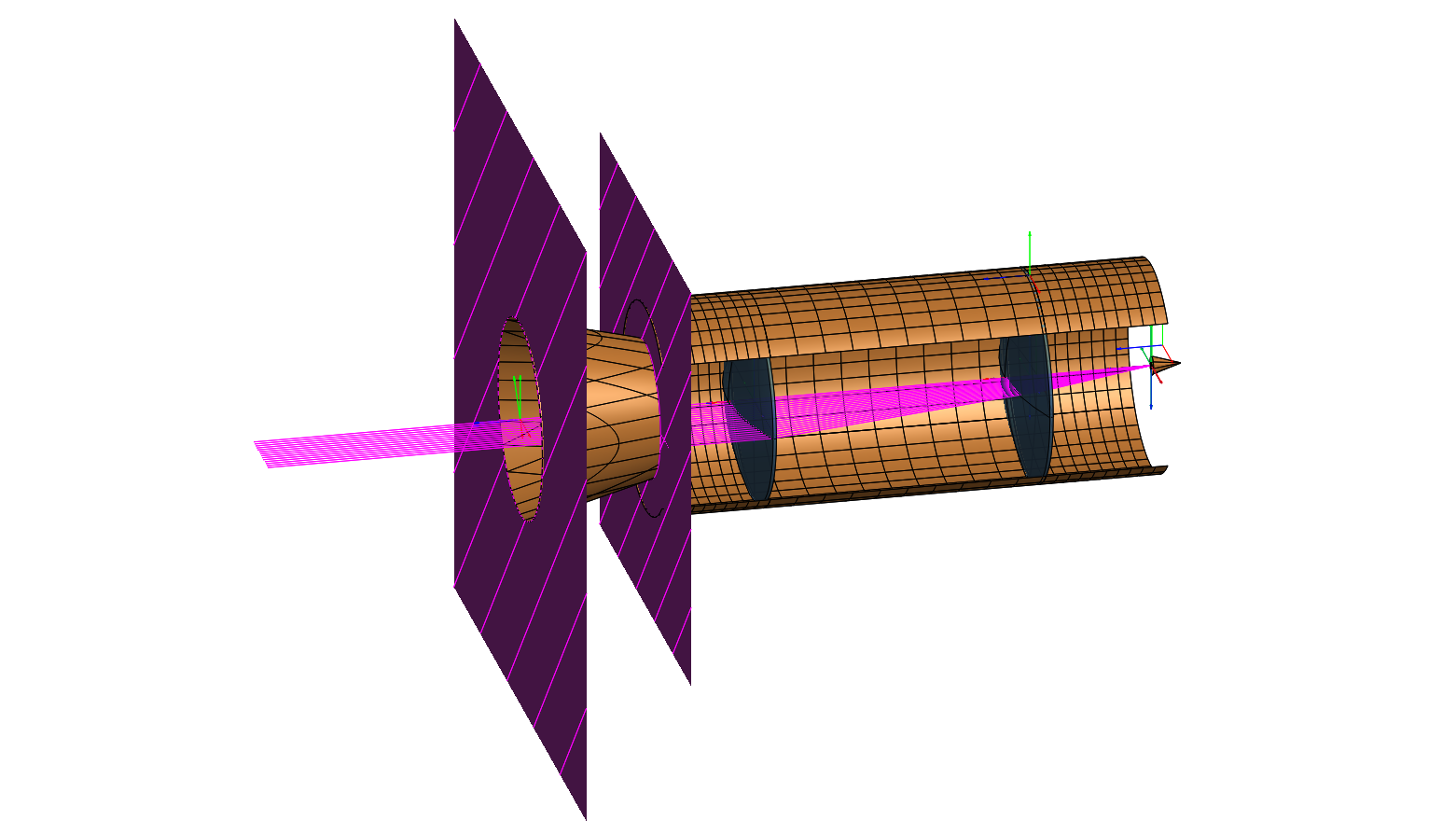}
    \end{minipage} &
    \begin{minipage}{0.4\hsize}
    \includegraphics[width = \hsize]{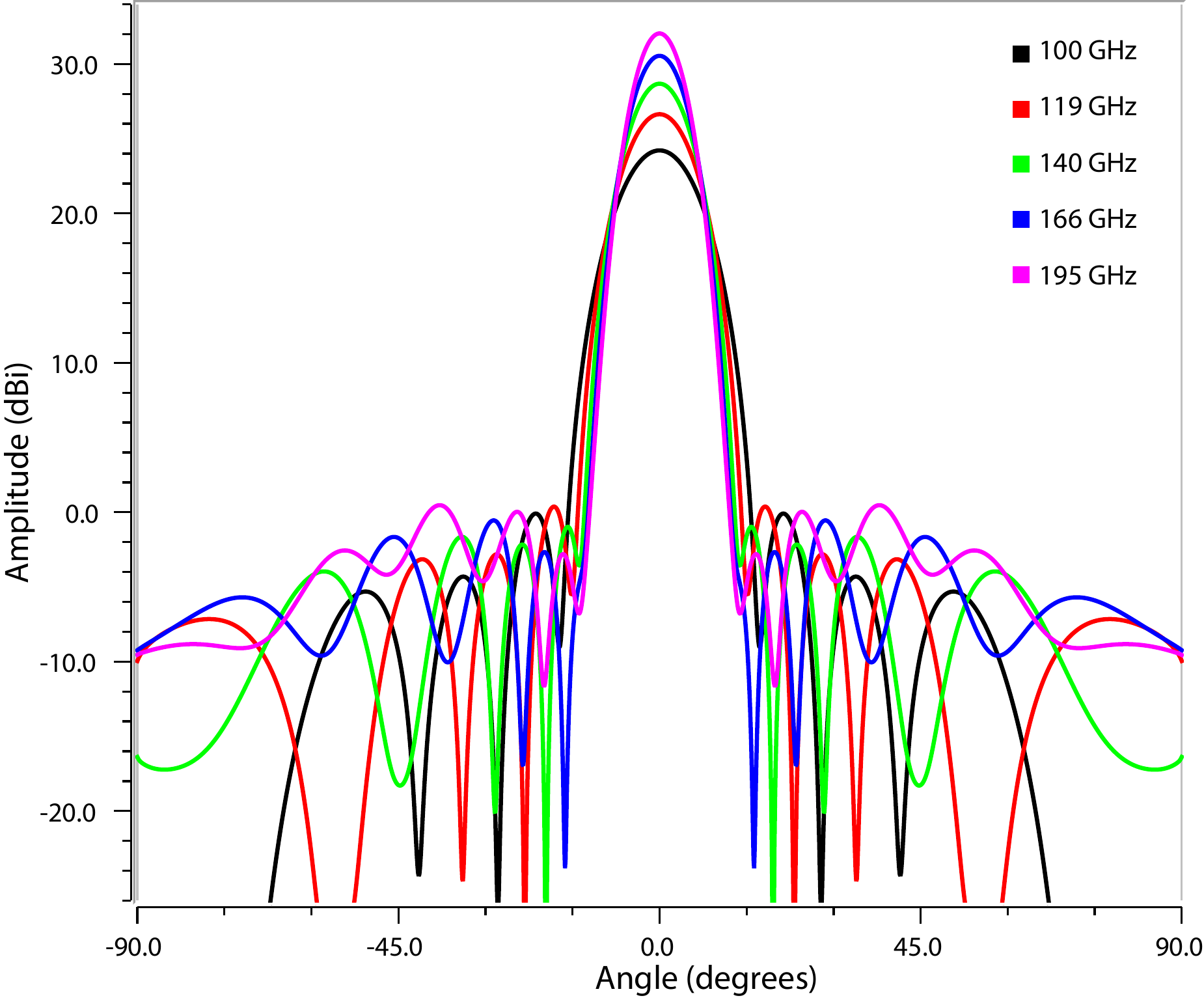}
    \end{minipage}
    \end{tabular}
    \caption{\CF{Left: MFT telescope model used to simulate the forebaffle flare angle impact on the MFT radiation patterns.  The telescope tube is not included in the simulation, and the forebaffle is assumed to be completely absorbing, so that only its aperture is considered. Right: the MFT beam former response is similar to that of a sinuous antenna coupled to a lenslet.}}
    \label{fig:FB_model}
\end{figure}

\CF{As an example, we now discuss the effect of the forebaffle size on the MFT beams. We simulated the MFT optics with physical optics, including the feed, the two lenses, the cold stop aperture and a fully absorbing forebaffle, as shown in Figure \ref{fig:FB_model}. Keeping the total length of the forebaffle fixed to its nominal value of 270 mm, we varied the flare angle within the range of 10° to 18° (the entrance aperture of the forebaffle is modified accordingly between 395 and 475 mm) to verify the impact on the radiation pattern at the MFT frequencies of 100, 119, 140, 166 and 195 GHz.}

\begin{figure}[t]
    \centering
    \begin{tabular}{cc}
    \includegraphics[width = 0.6\textwidth]{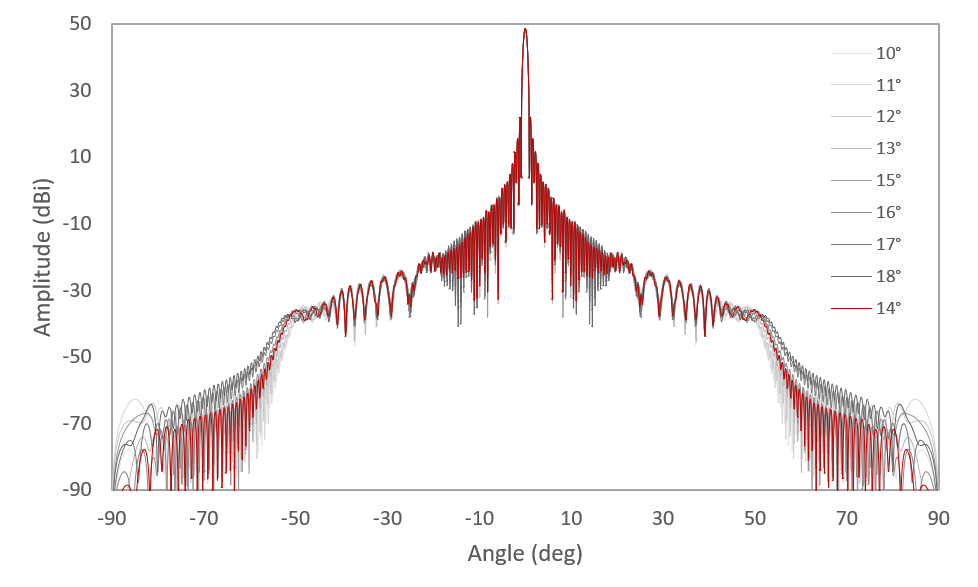}
    \end{tabular}
    \caption{\CF{Radiation pattern at 100 GHz for the MFT on-axis pixel with varying forebaffle flare angle.}}
    \label{fig:FB_sims}
\end{figure}

\CF{Figure \ref{fig:FB_sims} depicts the co-polar beams (phi = 90°) at 100 GHz for the on-axis pixel. Lower flare angles determine an increase of order of 3.5 dB in the intermediate sidelobe levels in the angular region between 10 and 15 degrees and a substantial increase of the far sidelobes at angles larger than $\pm50$ degrees, still about 80 dB below the main directivity.}

\CF{The entire analysis includes the co-polar and cross-polar response for on-axis and off-axis pixels on the focal plane. Similar analyses will be performed for the different elements of the MHFT telescopes in order to assess their impact on the final beam, also by combining those components which have a non-negligible impact on the optical response.}

\subsection{Feed simulations}
The feed beam patterns have been redundantly simulated with mode-matching and full 3D electromagnetic modeling software. After validating the basic results, the full-wave software Ansys HFSS has been used to generate a first set of simulations, thus providing lenslet beam patterns at $89$--$112$\,GHz for MFT and feedhorn beam patterns in the $402$\,GHz band of HFT. These feed patterns have been used so far for preliminary studies of spillover effects. Work is underway to include these patterns in a full telescope simulation to inform TOD-based simulations with more realistic beams.
A longer-term project involves therefore the production of the full frequency-dependent set of lenslet and feed  patterns. Such activity is expected to be particularly useful for the HFT, for which a full system-level numerical simulation is well out of reach of the available modeling/computing resources.  

\subsubsection*{MFT}
\begin{figure}[htb]
    \centering
    \begin{tabular}{cc}
    \begin{minipage}{0.48\hsize}
    \centering
    \includegraphics[width=1.2\hsize]{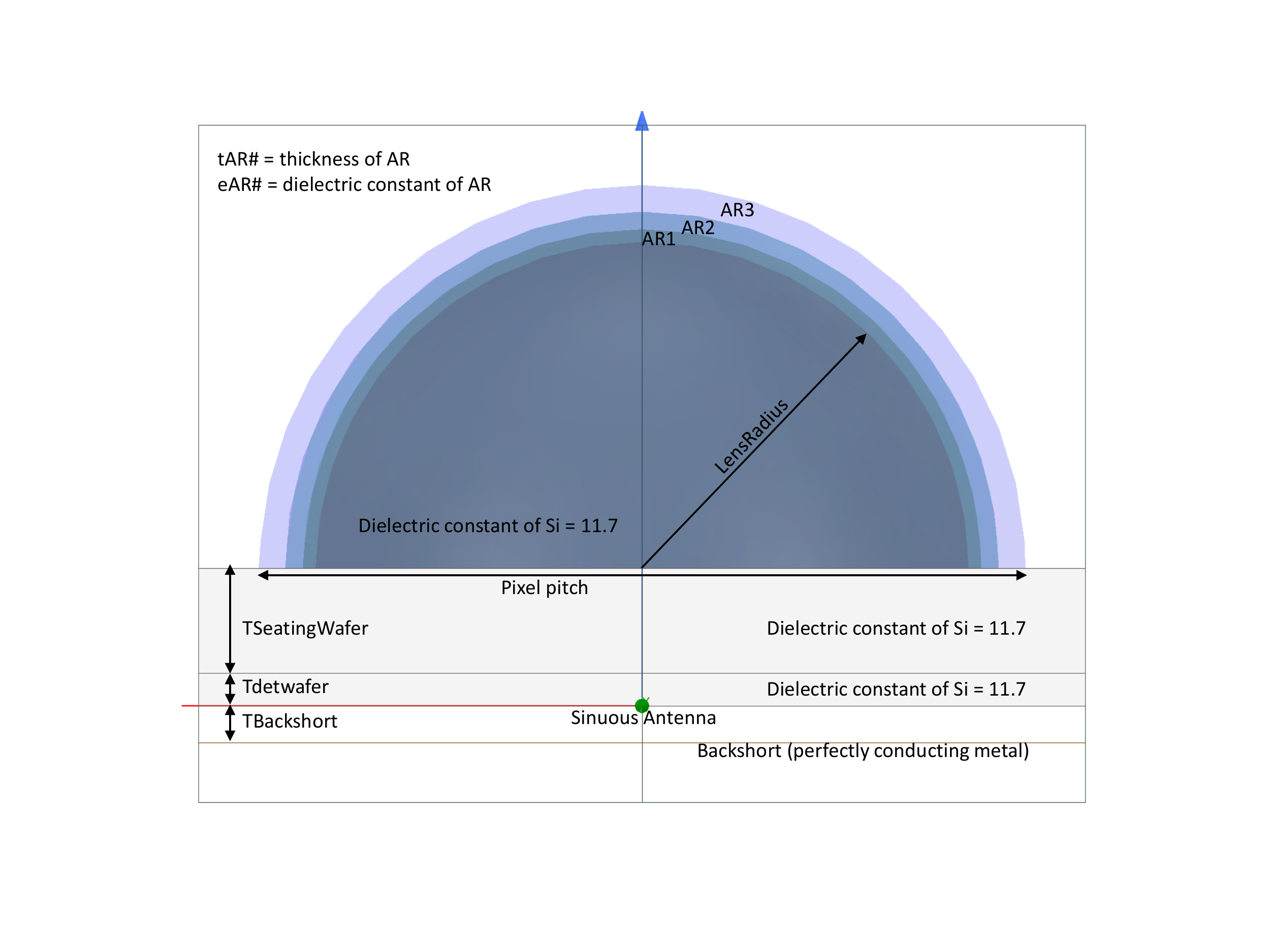}
    \end{minipage}&
    \begin{minipage}{0.45\hsize}
    \centering
    \includegraphics[width=\hsize]{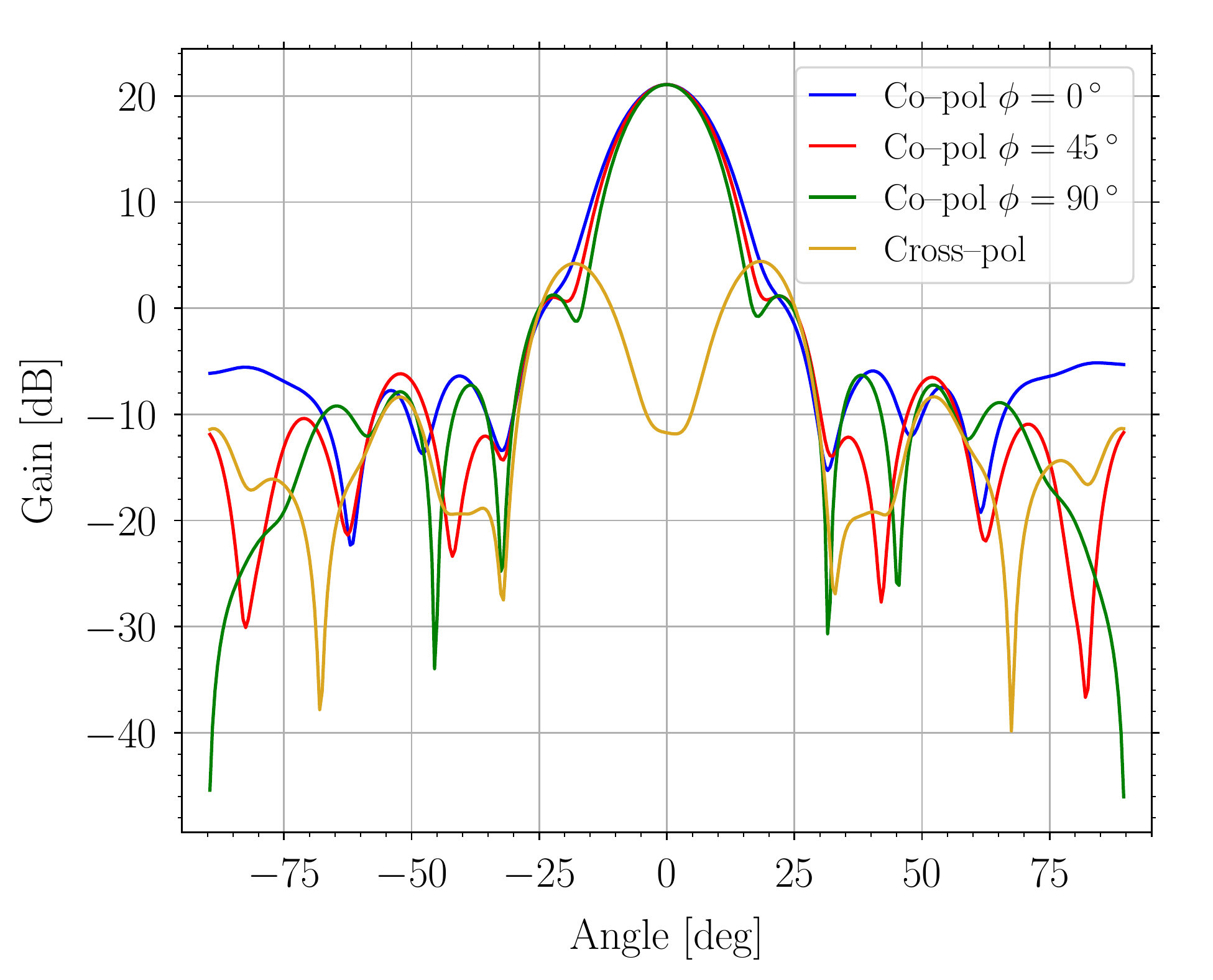}
    \end{minipage}
    \end{tabular}
    \caption{{\it Left:} \textcolor{red}{} sketch of baseline MFT coupling devices, made of a sinuous antenna and AR-coated silicon lenslet {\it Right:} Beam cuts at $100$\,GHz. Copolar beam cuts are shown at $\phi=0^{\circ}$, $45^{\circ}$ and $90^{\circ}$. Maximum cross-polarization is shown for $\phi=45^{\circ}$.}
    \label{fig:mft_lenslet}
\end{figure}

The MFT focal plane will feature sinuous antenna-coupled detectors with a silicon AR-coated lenslet\cite{Suzuki2018,Jaehnig2020}. The lenslet diameter is 12\,mm for all the MFT bands. A sketch of the device and the corresponding simulated beam patterns are shown in Fig.~\ref{fig:mft_lenslet}.

The MFT detector array consists of seven hexagonal wafers, each of which holds 61 pixels. To observe multiple bands, band-pass filters and multiple TES detectors are built into multichroic assemblies. The total number of the detectors in MFT is 2074.
\subsubsection*{HFT}
\begin{figure}[htb]
    \centering
    \begin{tabular}{cc}
    \begin{minipage}{0.48\hsize}
    \includegraphics[width=1.05\hsize]{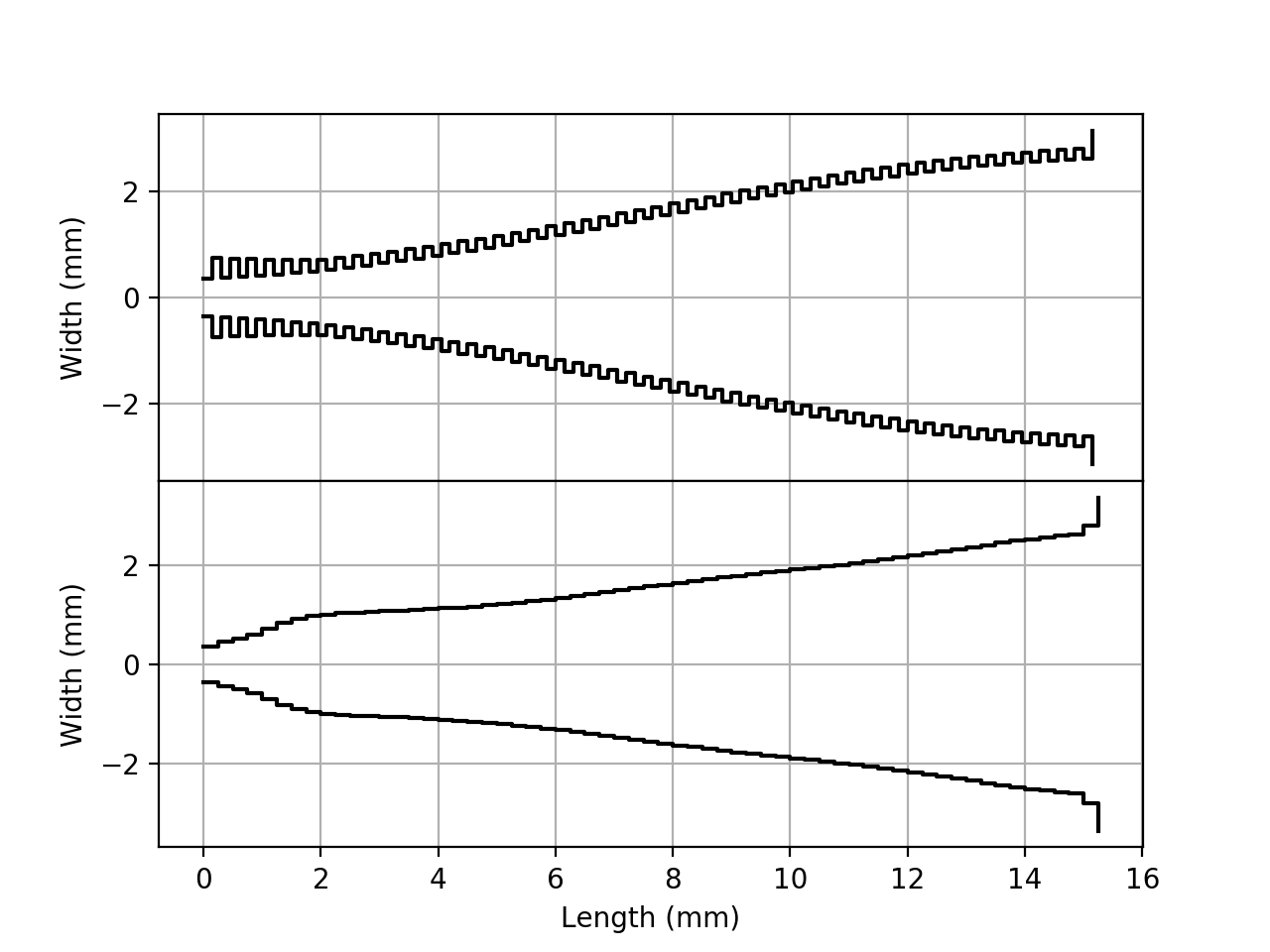}
    \end{minipage}&
    \begin{minipage}{0.47\hsize}
    \vspace{0.8cm}
    \includegraphics[width=\hsize]{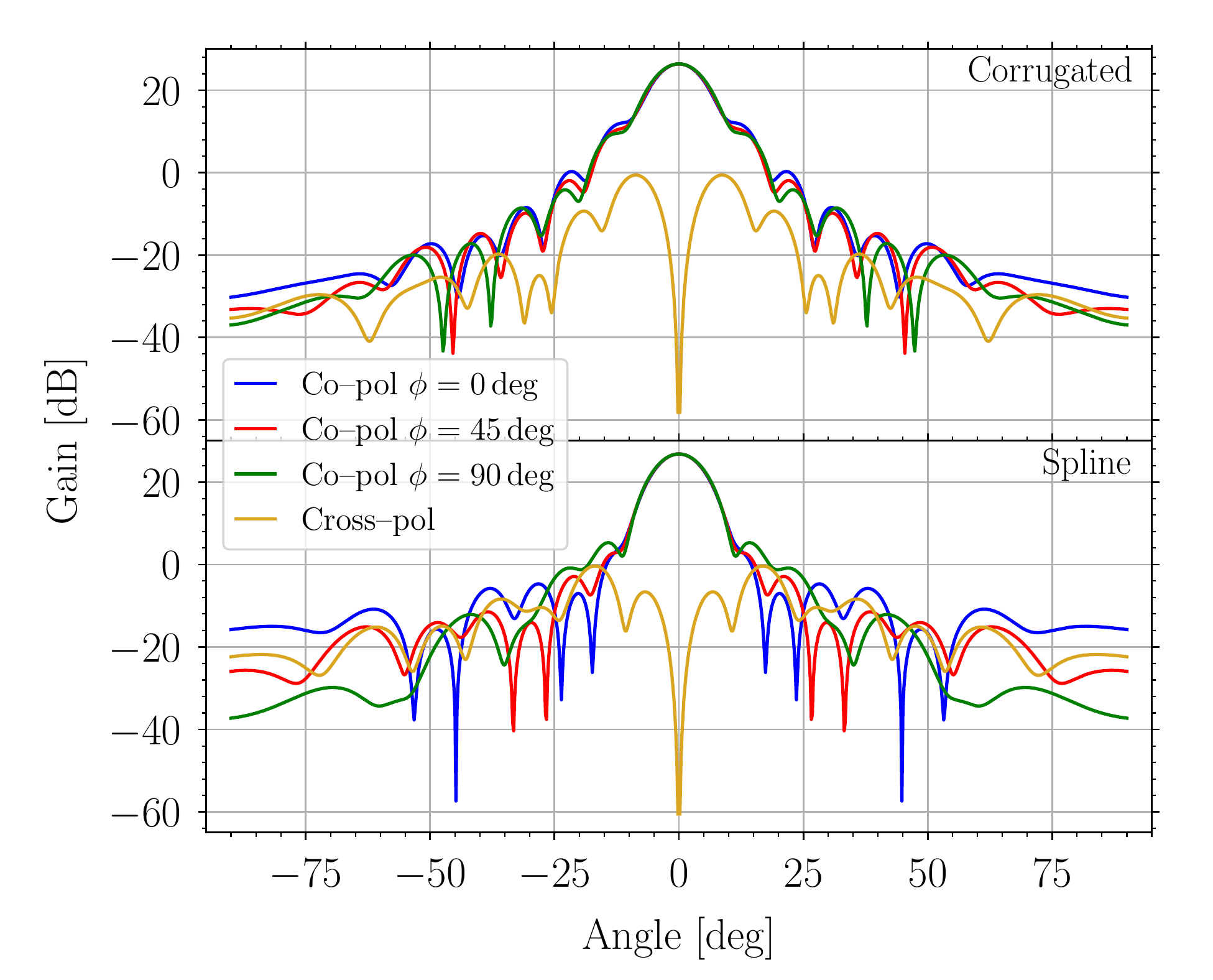}
    \end{minipage}
    \end{tabular}
    
    \caption{ {\it Left:} Corrugated and spline-profiled feedhorns proposed for the HFT highest frequency baseline channel. {\it Right:} Beam cuts of candidate HFT feedhorns at 450GHz. Copolar beam cuts are shown for $\phi=0^{\circ}$, $45^{\circ}$ and $90^{\circ}$. Maximum cross-polarization is shown for $\phi=45^{\circ}$. }    
    \label{fig:hft_horn}
\end{figure}

The HFT will employ horn-coupled detectors\cite{Suzuki2018}. The horn diameters are 6.6\,mm for the 195 to 338-GHz bands and 5.7\,mm 
for the 402 GHz band.
The 6.6-mm horns make two 127-pixel arrays with planar Ortho-Mode Transducers (OMTs).
Each 6.6-mm horn array can observe two bands and two linear polarizations, simultaneously, so that 254 TESs are employed.
The 5.7-mm horns make a 168 pixel monochromatic array with OMT. The total number of the detectors in HFT is 1354.
Two different horn profiles have been proposed for the  402 GHz detectors. Both are based on Si platelet technology \cite{nibarger2012} and are shown in the left panel of Fig.~\ref{fig:hft_horn}. The first profile is a corrugated feedhorn made of $101$ $150\mu m$--thick platelets, while the second is a spline-profile horn, adapted to the platelet technology in order to be made of $61$ $250\mu m$--thick platelets. Nominal performance for both models is excellent, as summarized in Tab.~\ref{tab:hft_horns}.
\begin{table}[htb]
    \centering
    \begin{tabular}{l|c|c}
    \hline
Horn type & Corrugated & Spline-profiled\\
Spillover efficiency avg (max) &$0.924$ $(0.933)$&$0.968$ $(0.986)$\\
Co-polar beam asymmetry avg (max)&$0.0037$ $(0.057)$&$0.0017$ $(-0.053)$\\
Cross-polarization avg (max) &$3.6\cdot 10^{-4}$ $(4\cdot 10^{-3})$& $13\cdot 10^{-4}$ $(6\cdot 10^{-3})$\\
Reflection loss avg (max)&$-31$dB $(-25$dB$)$&$23$dB $(-17$dB$)$\\
\hline

\end{tabular}
\caption{Synthetic data for proposed HFT horn geometries. Beam-integrated quantities refer to a nominal coupling F-number of 2.2. Maximum and average values refer to the highest HFT frequency channel, with a center frequency of 402\,GHz and a 25\% bandwidth.}
\label{tab:hft_horns}
\end{table}

\begin{figure}[htb]
    \centering
    \begin{tabular}{cc}
    \begin{minipage}{0.48\hsize}
    \includegraphics[width=\hsize]{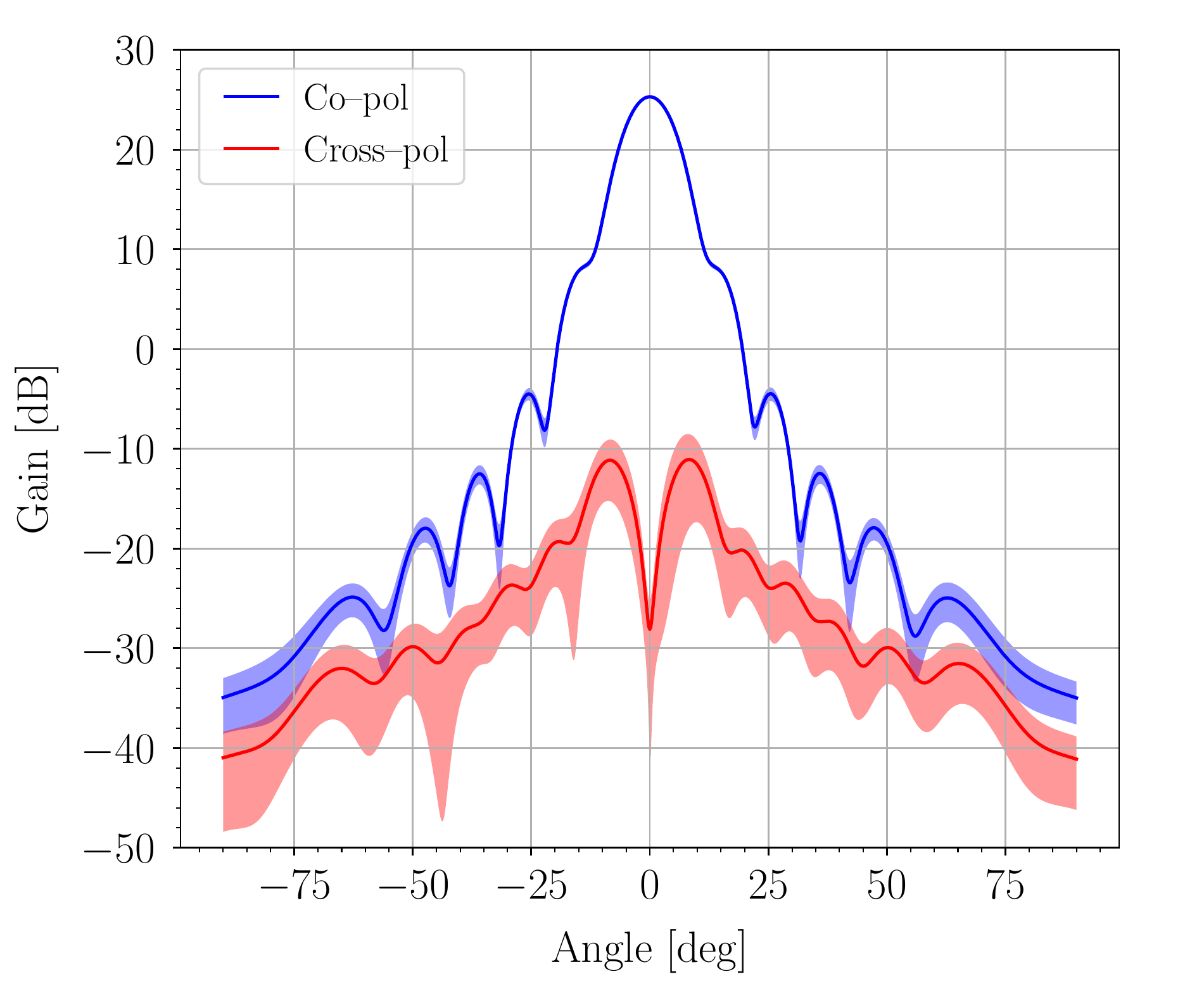}
    \end{minipage}&
    \begin{minipage}{0.48\hsize}
    \includegraphics[width=\hsize]{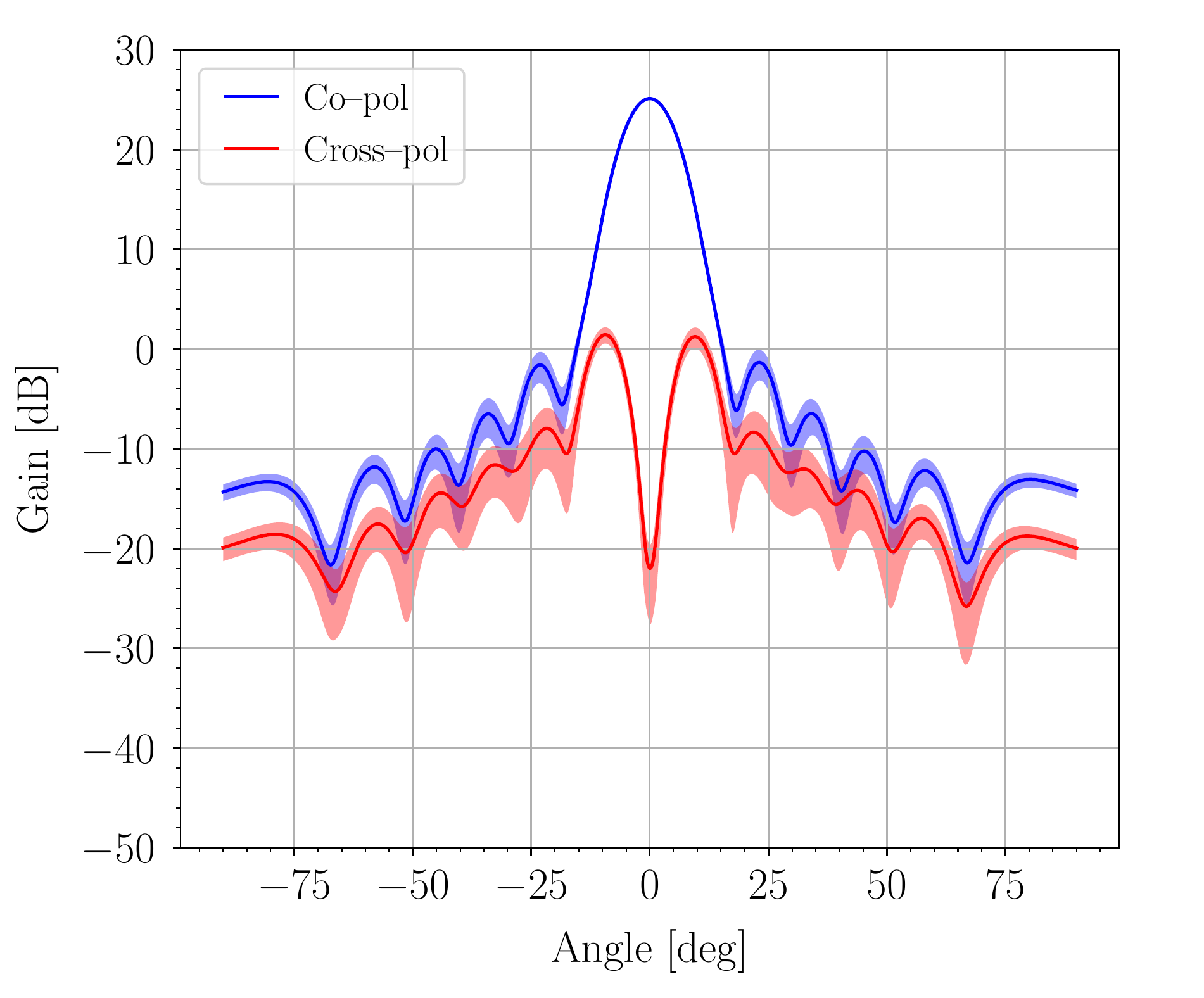}
    \end{minipage}
    \end{tabular}
    \caption{Beam cuts of candidate HFT feedhorns for random, $3\mu m\ rms$ platelet misplacements. Copolar beam cuts are shown for $\phi=0^{\circ}$. Cross-polarization is shown for $\phi=45^{\circ}$. {\it Left:} Corrugated feedhorn. {\it Right:} Spline-profiled feedhorn. Colored bands highlight the region containing 68\% of realizations.}    
    \label{fig:hft_tol}
\end{figure}

In order to check the consistency of current fabrication and assembly tolerances of silicon platelet feedhorns with the global coupling requirements of the HFT, an analysis has been started to investigate the impact of plate-to-plate misalignment on beam properties. 
Fig.~\ref{fig:hft_tol} shows that the copolar and cross-polar beam patterns are very stable for both horn designs with $3\mu m$ rms displacements between neighboring platelets. This level of alignment has been achieved in previous silicon platelet horn builds by use of a simple pin and slot alignment method.   
The band-averaged HFT spillover efficiency of the corrugated (spline-profiled) horn is $\sim93\%$ ($\sim96\%$), and changes by no more than $2\%$ over $100$ misalignment realizations for either profile.  
We also find that the peak cross-polar response is $<0.7\%$ for all realizations.
These results hold up to the upper edge of the highest HFT band, suggesting that both designs are suitable for HFT implementation. Additional simulation work is underway to investigate more extreme or self-correlated misplacement profiles along the horn.


\section{Modeling Challenges and mitigating actions}
\label{sec:challenges}
\subsection*{Optical Ghosts}
\begin{figure}
    \centering
    \begin{tabular}{cc}
    \begin{minipage}{0.34\hsize}
    \includegraphics[width=\hsize]{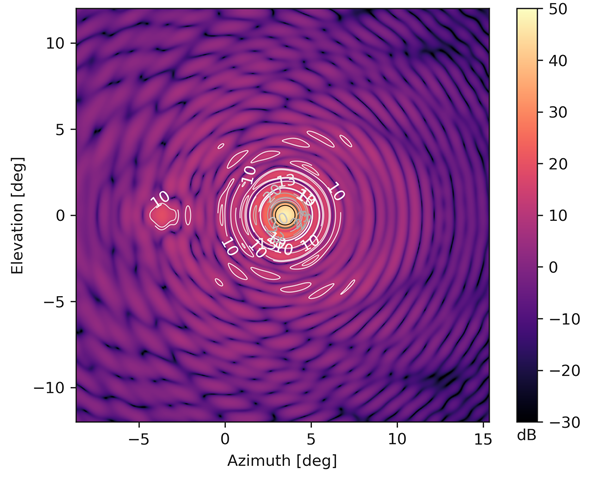}
    \end{minipage}&
    \begin{minipage}{0.61\hsize}
    \includegraphics[width=\hsize]{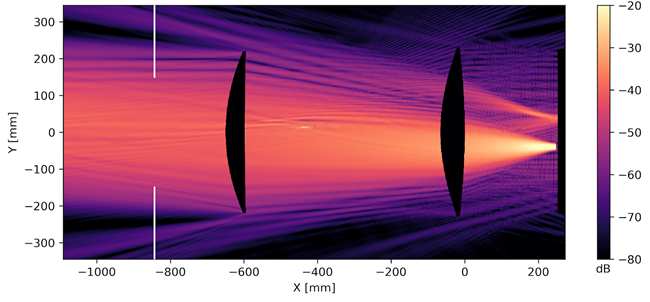}
    \end{minipage}
    \end{tabular}
    \caption{ {\it Left:} Optical ghosting at 90\,GHz resulting from the combined effect of non-ideal ARC on MFT lenses and imperfect absorption on the focal plane, as seen in the telescope far field. Radiation is launched from a focal plane positon corresponding to a 40\,mm off-center detector. {\it Right:} Intensity distribution in the telescope plane for the same combination of effects.}    
    \label{fig:ghost_examples}
\end{figure}

\LL{Telescope ghosts are position-sensitive artifacts originating by field propagation across the optical system, due to  non-ideal behavior of the optical elements and of supporting/surrounding structures. A few examples are an imperfect absorption by tube walls, non-ideal AR coatings on HWP and lens surfaces, with consequent frequency and polarization dependent reflection and scattering, and an imperfectly matched focal plane coupling. Radiation re-launched from these effects on optical paths getting it at least partially refocused on the focal plane will be detected as more or less diffused structures superimposed over the actual image of the sky, potentially yielding systematic effects in signal recovery.
Preliminary physical optics simulations have been run to check the ability of common simulation algorithms to capture some of these effects. As an example, Fig.~\ref{fig:ghost_examples} shows the effects of non-ideal single-layer ARCs on the 90\,GHz beam pattern and field distribution in a simplified version of the MFT, for a nominal pixel position 40\,mm off the center of the focal plane.}

\LL{The effort towards a robust modeling and understanding of the relative contributions of ghosts to the systematic effects is still ongoing. 
Scattering effects, in particular, are not easily captured in existing software codes, especially when system level simulations are performed.
This is why a validation campaign based on experimental measurements on real-life sample systems is needed to ensure a correct evaluation of the ghosting problem across the whole MFT and HFT frequency range.}


\subsection*{Polarization Modulator}
{\AR The continuous rotation by the HWP of the input polarization provides quasi-simultaneous observations  of the I, Q and U Stokes parameter across the focal plane pixels. In addition, thanks to the HWP modulation, the input polarization is shifted at higher  frequency with reference to the slower signal fluctuations that are usually associated with any kind of noise coming from the instrumentation: electronics, cryogenics etc. 

However, imperfections of the HWP lead to the modulation of the background signal which appears to peak at harmonics of the HWP rotation frequency in the data spectrum. This kind of instrumental noise is known as HWP parasitic signal (HWPSS) and has been observed  in previous experiments, such as Maxima \cite{Johnson_2007}, EBEX \cite{chapman2014}, NIKA \cite{ritacco2017}. Methods are available to model this signal and subtract it in the data analysis step from the TOD of each detector \cite{ritacco2017}. 

In addition, any departure from a Gaussian shape of the beam can cause  leakage of the total intensity to polarization, resulting in a beam pattern deformed in polarization \cite{ritacco2017}. This effect might be observed in the Stokes Q and U maps of an unpolarized source and we refer to it as instrumental polarization. The modelling of this beam pattern is usually more difficult \cite{2020EPJWC.22800002A} for large arrays of detectors because each detector might have a slightly different beam. For this reason the knowledge of the beams and the study of the effect through simulations and laboratory tests with the full optics, including the HWP, is crucial to constrain such a systematic in the final data.}

\subsection*{Absorbers}
Quasi-optical filters will be designed and implemented to reject unwanted radiation propagating from the sky all the way down to the focal plane. Nonetheless, in-band radiation may picked up as a consequence of scattering off or re-emitted from optical elements or devices along the optical chain. As already anticipated, this is responsible for excess loading and/or image artifacts on the focal planes. The use of absorbing coatings and materials, possibly complemented by a careful shaping of the internal surfaces of the tube assemblies, is a standard technique to mitigate such issues. 
Ideally, a perfect absorber emits an exact blackbody spectrum (in which case the loading on the focal planes depends only on the absorber temperature), absorbs efficiently radiation well away from normal incidence, and features a large thermal inertia, so that the radiating/absorbing properties don't exhibit thermal drifts which might ultimately determine dangerous time-constant effects in the detectors response.
In addition, any candidate absorber for the MHFT tubes must comply with the tight mass allocation constraints and meet the requirements for space qualification. 

A wealth of space-qualified absorbers have been proposed in the past for far-infrared missions\cite{Persky1999,baccichet2015}. These are mostly based on dielectric molds loaded with conductive scattering powders, and their characteristics are available in the literature down to THz frequencies. Other materials, like Eccosorb CR-110, have long been known to approach blackbody emissivity down to tens of GHz\cite{Halpern1986}. Recent work has been devoted to characterize loaded epoxy powders as absorbers at THz and sub-THz frequencies\cite{Wollack2008,Zivkovic2011,Ghigna2020}.
Metamaterial-based absorbers, based on low-frequency adaptations of THz metamaterial designs\cite{Chen2015} have also been recently proposed, even if their readiness for space environment and actual applicability on non-flat surfaces have yet to be explored.
3D-printed absorbing structures have also been recently demonstrated in selected bands below 230\,GHz\cite{Petroff2019}.

Regardless of the technology of the adopted solution, MHFT absorbers need to be carefully measured and modeled to assess their behavior at low temperature (the MFT and HFT tubes operate at 5K) in terms of incidence-, polarization- and frequency-dependent reflectance and scattering properties. Most of these effects cannot be easily implemented in numerical simulations and their treatment must rely on experimental data for the specific materials involved at the MHFT frequencies. 
Absorbers around the cold aperture stop, in particular, are a source of major concern because of the close vicinity to the warm PMU mechanism. They are meant to mitigate the excess radiation leaking in the tube through diffraction and scattering off the edges of the aperture stop. 

The current effort for MFT and HFT absorber studies is twofold: 
\begin{itemize}
    \item Representative subsystem-level simulations are performed to optimize surface shaping as a function of coating thickness, typical illumination and likelihood of unwanted pickup. 
    \item A systematic characterization campaign has been started to update the available information on known materials or molds  thorugh dedicated measurements on samples: spectral/polarimetric room-temperature measurements of reflectance and scattering will be followed by representative tests in cold environment. 
\end{itemize}
The results of these studies are meant to inform the tradeoff between electromagnetic performance and impact on the mass budget, and to inspect the need for a passive or active thermal control of the most sensitive regions of the tubes.

\section{Supporting laboratory measurements}
\begin{figure} [ht]
   \begin{center}
   \begin{tabular}{cc} 
   \includegraphics[width=0.4\hsize]{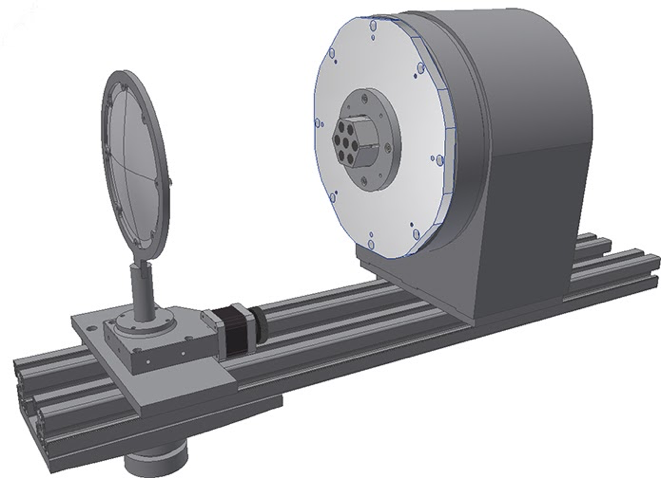}
   &
   \includegraphics[width=0.4\hsize]{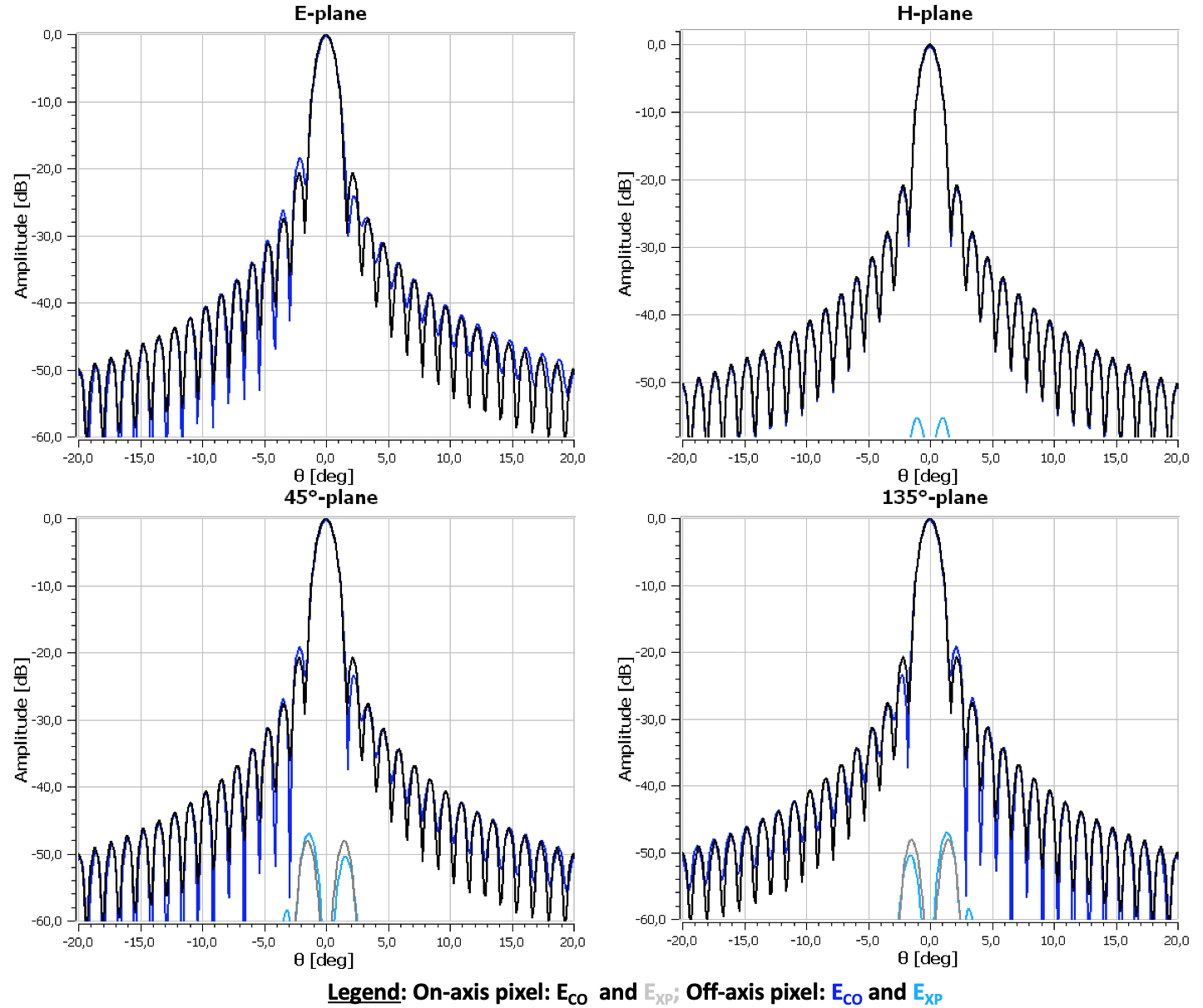}
   \end{tabular}
   \end{center}
   \caption{{\it Left: }BreadBoard 1 configuration. {\it Right: } Simulated far-field beam patterns corresponding to BB1. The figure shows co-polarization (E$_{CO}$) and cross-polarization (E$_{XP}$) field amplitude resulting from the illumination of the lens by on-axis and off-axis pixels for different azimuthal cuts (E- / H- / 45°-  /135°- planes).} 
\label{fig:bbm1} 
\end{figure}

{\CF The overall LiteBIRD requirements, dictated by the science goals, translate into a need in the knowledge of the optical system to unprecedented levels. We therefore need to rely extensively on dedicated experimental characterisation in order to verify our assumptions, build and consolidate our models. These experimental measurements need to happen at different levels, starting from material characterization to sub-system and system-level performance assessment. Similarly to any space mission, several models of the instruments will be developed over the years in order to build our knowledge of the overall performances of MHFT (typically prototype, qualification model followed by the flight model). However several fundamental questions need to be  answered very soon in order to gain confidence in the results of our modelling tools. Questions such as the knowledge of material optical characteristics at very low temperature, impact of potential in-homogeneity of the lens material, accuracy of refractive optics modelling, mainly when polarisation is concerned, still need to be answered to the required accuracy level. Therefore preliminary characterisations will be performed in parallel, on materials and on breadboard models, starting from a very simple configuration (BB1 - BreadBoard 1) evolving then to more complete and complex optical concepts.\\
BB1 comprises only one lens coupled to a small array of corrugated feedhorns (left panel of Fig.~\ref{fig:bbm1}), allowing to compare predicted and measured co- and cross-polarisation beams for on-axis and off-axis pixels. The first GRASP modelling results are presented in the right panel of Fig.~\ref{fig:bbm1}. Through this simple concept the accuracy of the modelling will be tested.\\
BB2 will then be a two lens system, again coupled to corrugated feedhorns in order to be closer to the MHFT optical design, even if not the same. In addition to improving the accuracy in the modelling, other tests such as positioning tolerances between optical components will performed.\\
While these breadboards will be characterised over the MHFT frequency range, cryogenics RF characterisation of materials, such as the ones for lenses, or again potential absorbers will be run in parallel in order to measure their properties at very low temperature. These tests will rely on Fourier Transform Spectrometer and quasi optical setups coupled to vector network analysers measurement systems.}


\section{Concluding remarks.}
We have provided an overview of the major optical studies and outstanding issues related to the Litebird MFT and HFT design. After describing the general issues of the optical design in relation to the unprecedented performance requirements of LiteBIRD, and in particular to the use of continuously rotating PMUs, we have described the baseline optical parameters along with the main design drivers which helped define the current optical configuration of MHFT.
We have shown preliminary analysis of the optical configuration for both telescopes and basic physical optics studies of the telescope properties, including tolerancing studies, forebaffle and spillover effects. Realistic feed simulations have been separately run to inform the physical optics studies with more realistic representations of the FPU and telescope coupling. Sample studies of higher-level effects, related to the presence of non-ideal HWPs, optical ghosts, imperfectly absorptive structures have been discussed.
We also pointed out the existence of a coordinated effort towards experimental validation of the adopted modeling tools and methods, in view of the next generation of optical studies, aimed both at final system design and optimization, and to the achievement of a more complete and realistic picture of instrument performance for end-to-end evaluation of scientific performance.

\acknowledgments 
This work is supported in \textbf{Japan} by \mbox{ISAS/JAXA} for Pre-Phase A2 studies, by the acceleration program of JAXA research and development directorate, by the World Premier International Research Center Initiative (WPI) of MEXT, by the JSPS Core-to-Core Program of A. Advanced Research Networks, and by JSPS KAKENHI Grant Numbers JP15H05891, JP17H01115, and JP17H01125. The \textbf{Italian} LiteBIRD phase A contribution is supported by the Italian Space Agency (ASI Grants No.~2020-9-HH.0 and 2016-24-H.1-2018), the National Institute for Nuclear Physics (INFN) and the National Institute for Astrophysics (INAF). The \textbf{French} LiteBIRD phase A contribution is supported by the Centre National d’Etudes Spatiale (CNES), by the Centre National de la Recherche Scientifique (CNRS), and by the Commissariat à l’Energie Atomique (CEA). The \textbf{Canadian} contribution is supported by the Canadian Space Agency. The \textbf{US} contribution is supported by NASA grant No.~80NSSC18K0132. 
\textbf{Norwegian} participation in LiteBIRD is supported by the Research Council of Norway (Grant No.~263011). The \textbf{Spanish} LiteBIRD phase A contribution is supported by the Spanish Agencia Estatal de Investigación (AEI), project refs. PID2019-110610RB-C21 and AYA2017-84185-P. Funds that support contributions from \textbf{Sweden} come from the Swedish National Space Agency (SNSA/Rymdstyrelsen) and the Swedish Research Council (Reg. no.~2019-03959). The \textbf{German} participation in LiteBIRD is supported in part by the Excellence Cluster ORIGINS, which is funded by the Deutsche Forschungsgemeinschaft (DFG, German Research Foundation) under Germany’s Excellence Strategy (Grant No.~EXC-2094~-~390783311). This research used resources of the Central Computing System owned and operated by the \textbf{Computing Research Center at KEK}, as well as resources of the \textbf{National Energy Research Scientific Computing Center}, a DOE Office of Science User Facility supported by the Office of Science of the U.S. Department of Energy.
The European Space Agency (\textbf{ESA}) has led a Concurrent Design Facility study, focused on the MHFT and Sub-Kelvin coolers, and funded Technology Research Programmes for “Large radii Half-Wave Plate (HWP) development” (contract number: \mbox{4000123266/18/NL/AF}) and for the ‘Development of Large Anti‐Reflection Coated Lenses for Passive (Sub)Millimeter‐Wave Science Instruments” (contract number: \mbox{4000128517/19/NL/AS}).

In addition, the authors would like to acknowledge the Ile de France region through its DIM-ACAV programme, the university Paris-Saclay through a chaire d’excellence, and the OSUPS that have supported the work presented here.
\bibliography{report} 
\bibliographystyle{spiebib} 

\end{document}